\documentclass[12pt]{article}
\usepackage{latexsym}
\font\tenmsbm=msbm10 scaled 1200
\font\sevenmsbm=msbm9
\newfam\msbmfam
\textfont\msbmfam=\tenmsbm
\scriptfont\msbmfam=\sevenmsbm

\def\msbm{\fam\msbmfam\tenmsbm}

\makeatletter
\@addtoreset{equation}{section}
\makeatother
\renewcommand{\theequation}{\thesection.\arabic{equation}}
\newcounter{parentequation}% Counter for ``parent equation''.
\newenvironment{subequations}{%
  \refstepcounter{equation}%
  \begingroup % conservative approach
\let\protect\noexpand
  \edef\@tempa{\def\noexpand\theparentequation{\theequation}}%
  \expandafter
  \endgroup\@tempa
  \setcounter{parentequation}{\value{equation}}%
  \setcounter{equation}{0}%
  \def\theequation{\theparentequation\alph{equation}}%
  \ignorespaces
}{%
  \setcounter{equation}{\value{parentequation}}%
}
\newcommand{\eqn}[1]{(\ref{#1})}
\newsavebox{\uuunit}
\sbox{\uuunit}
    {\setlength{\unitlength}{0.825em}
     \begin{picture}(0.6,0.7)
        \thinlines
        \put(0,0){\line(1,0){0.5}}
        \put(0.15,0){\line(0,1){0.7}}
        \put(0.35,0){\line(0,1){0.8}}
       \multiput(0.3,0.8)(-0.04,-0.02){10}{\rule{0.5pt}{0.5pt}}
     \end {picture}}
\newcommand {\unity}{\mathord{\!\usebox{\uuunit}}}
\newsavebox{\bobox}
\sbox{\bobox}
    {\setlength{\unitlength}{0.825em}
     \begin{picture}(0.6,0.7)
        \thinlines
        \put(0,0){\line(1,0){0.7}}
        \put(0,0){\line(0,1){0.7}}
        \put(0.7,0){\line(0,1){0.7}}
        \put(0,0.7){\line(1,0){0.7}}
        \multiput(0,0)(0.05,0.05){14}{\rule{0.4pt}{0.4pt}}
        \multiput(0,0.65)(0.05,-0.05){13}{\rule{0.4pt}{0.4pt}}
     \end {picture}}
\newcommand {\boxtimes}{\mathord{\!\usebox{\bobox}}\,}
\def\IC{\relax\,\hbox{$\inbar\kern-.3em{\rm C}$}}
\def\bfzFero{\relax\,\hbox{$\inbar\kern-.3em{\rm 0}$}}
\def\IR{\hbox{\msbm R}}

\def\mD{\hbox{I$\!$D}}
\def\bfone{\relax{\rm 1\kern-.35em 1}}
 \def\cB{{\cal B}}
 \def\cD{{\cal D}}  
 
\def\cH{{\cal H}}

\def\cN{{\cal N}}

\def\beq{\begin{equation}}
\def\eeq{\end{equation}}
\def\bea{\begin{eqnarray}}
\def\eea{\end{eqnarray}}
\def\bet{\begin{tabular}}
\def\eet{\end{tabular}}
\def\bes{\begin{subequations}\bea}
\def\ees{\eea\end{subequations}}
\newcommand{\dfrac}{\displaystyle \frac}

\def\a{\alpha}
\def\b{\beta}

\def\g{\gamma}

\def\s{\sigma}
\def\e{\epsilon}

\def\l{\lambda}
\def\hT{\hat{T}}
\def\IC{\hbox{\msbm C}}
\begin{document}

\begin{flushright}
DFTT 40/99
\end{flushright}

\vspace{1cm}

\begin{center}

{\Large \bf KK Spectroscopy 
of Type IIB Supergravity on $AdS_5 \times T^{11}$}

\vspace{2cm}

{ Anna Ceresole$^{\star}$}\footnote{ceresole@athena.polito.it},
{Gianguido Dall'Agata$^{\dag}$}\footnote{dallagat@to.infn.it},
and
{Riccardo D'Auria$^{\star}$}\footnote{dauria@polito.it}

\vspace{1cm}

{ $\star$ \it Dipartimento di Fisica, Politecnico di Torino  \\
C.so Duca degli Abruzzi, 24, I-10129 Torino, and\\
Istituto Nazionale di Fisica Nucleare, Sezione di Torino. \\
}

\medskip

{$\dag$ \it Dipartimento di Fisica Teorica, Universit\`a  di Torino 
and \\
Istituto Nazionale di Fisica Nucleare, Sezione di Torino, \\
 via P. Giuria 1, I-10125 Torino.}

\end{center}

\vspace{1cm}

\begin{abstract}
We give full details for the  computation of the Kaluza--Klein mass spectrum
of Type IIB Supergravity on $AdS_5 \times T^{11}$,  with $T^{11}=SU(2)\times
SU(2)/U(1)$, that has recently lead to both stringent tests and interesting
predictions on the $AdS_5/CFT_4$ correspondence for $\cN=1$ SCFT's \cite{CDD}.
We exhaustively explain how KK states arrange into  $SU(2,2|1)$
supermultiplets, and stress some relevant features of the $T^{11}$ manifold,
such as  the presence of topological modes in the  spectrum originating from
the existence of non--trivial 3--cycles.  The corresponding Betti vector
multiplet is responsible for the extra baryonic symmetry in the boundary $CFT$. 
More generally, we use the simple $T^{11}$ coset  as a laboratory to revive
the technique and show  the power of  KK harmonic expansion, in view of the
present attempts to probe along the same lines also $M$--theory 
compactifications and the $AdS_4/CFT_3$ map.
\end{abstract}

\newpage

\baselineskip 6 mm

%%%%%%%%%%%%%%%%%%%%%%%%%%%%%%%%%%%%%%%%%%%%%%%%%%%%%%%%%%%%%%%%%%%
%%%%%%%%%%%%%%%%%%%%%%%%%%%%%%%%%%%%%%%%%%%%%%%%%%%%%%%%%%%%%%%%%%%
\section{Introduction}

Knowledge of the full Kaluza--Klein (KK) mass spectrum of Type IIB 
Supergravity compactified on the product of $AdS_5$ spacetime and the 
coset manifold $T^{11}=\dfrac{SU(2)\times SU(2)}{U(1)}$ has recently 
been crucially used to perform accurate spectroscopic tests as weel as
obtain many new predictions \cite{CDD} on the $AdS/CFT$ 
correspondence \cite{M} by comparison with a specific $\cN=1$, 
four--dimensional SCFT \cite{KW}.

Although partial results concerning the $T^{11}$ laplacian \cite{G}, 
the first--order operators acting on fermions and on the $2$--form 
\cite{JR} were already quite inspiring, only from studying 
every sector of the spectrum one can fully analyse the multiplet 
structure and shortening patterns that allow to establish a precise 
mapping between KK states and conformal operators in the boundary 
field theory.

While the focus in \cite{CDD} was on the $AdS/CFT$ correspondence, 
this paper is totally devoted to the techniques involved in performing 
harmonic analysis on $T^{11}$, that were just touched in 
that previous work.

Harmonic analysis, applied to the KK reduction of higher supergravity
theories, has been thoroughly developed in the past (see
\cite{SS,Duffrev,libro,Sdd} and references therein).  Nowadays, it seems to
live a new youth, especially in connection with the basic r\^ole played by all
$AdS$ compactified supergravity, and particularly of $M$--theory .

The $T^{11}$ manifold provides a simple and effective example of coset 
manifold with Killing spinors that can be used to show the general 
theory of computing KK spectra at work.  Thus we briefly expose it 
here in the essential steps without any claim for mathematical rigour.  
This can perhaps make more evident points presented in \cite{CDD} and 
might be of some help to those who are not aware of the method.

Since we determine the full KK spectrum, we organize it in $SU(2,2|1)$ 
supermultiplets and discuss some interesting features arising in this 
kind of analysis, we think that this paper could provide a 
self--contained reference for analogous computations in different 
examples of KK harmonic expansion that are nowadays under 
investigation, such as $M$--theory on $AdS_4\times M_7$ \cite{tutti},
 with $M_7$ being either $M^{111}$ spaces \cite{M111,cdfpv,Betti2,Malt}, 
the real Stiefel manifold 
$V_{(5,2)}$ \cite{CDD3} or the $N^{010}$ and $Q^{111}$ spaces \cite{TG}.

The strength of our analysis lies in the fact that, in view of the 
$AdS/CFT$ correspondence, our results provide an accurate check of 
the field theory independently derived by orbifold resolution 
techniques \cite{KW,orb,MP}.
This is not the case for the $AdS_4/CFT_3$ correspondence, where 
the supergravity side can be deeply investigated and is partly well known, 
while the relevant three--dimensional conformal field theories are not 
still understood, although there is some work in this direction 
\cite{PS,OT,GD}.
The KK analisys acquires than a prominent position, since it can give 
us the right hints for building such dual CFT's.

Beside the harmonic analysis and the KK mass spectrum, we also give 
some emphasis to the existence on $T^{11}$ of a ``Betti" vector 
multiplet \cite{Betti2,DaF}, that is a gauge vector multiplet which 
is a singlet of the full isometry group $SU(2) \times SU(2) \times 
U_R(1)$ and whose presence is related to the non--trivial Betti 
numbers $b_2 = b_3 = 1$ on $T^{11}$.
As already observed in \cite{CDD,KW2}, the existence of such kind of 
multiplets is related to the possibility of wrapping a $p$--brane 
($p=3$ in our case) on the non--trivial $p$--cycles of the internal 
manifold.

A new feature of our five--dimensional theory is  that
there appears not only the single vector field associated to the non--trivial 
$3$--form, but, as we will see, there is also a tensor and a scalar 
field, which are part of a Betti tensor and a Betti hypermultiplet 
respectively.

In section 2 we review in some more detail with respect to previous 
papers the geometry of $T^{pq}$ and specifically  of $T^{11}$ (which 
turns out to be the only supersymmetric one in the family where the 
integers $p$ and $q$ define the possible embeddings of the $U(1)$ in 
the denominator), establishing conventions and notations.
In section 3 we give our essential resum\'e of  harmonic expansion 
on a general coset manifold, giving at each step of the construction, 
the  application to our specific manifold.
The KK procedure for $T^{11}$ is then worked out and the mass spectrum 
thoroughly computed in section 4.  
In section 5 we explain how to reconstruct the supermultiplets of 
$SU(2,2|1)$ from the eigenvalues of the invariant differential 
operators and from the group theoretical knowledge of the $SU(2,2|1)$ 
representations.
In section 6 we recall the main properties of the Betti multiplets
as they were first introduced in \cite{DaF} and we discuss their
explicit form in our specific case.

We also present three appendices.  In the Appendix A notations and 
conventions are collected, while in Appendix B we show how to compute 
the differential operators on a coset manifold in a purely algebraic 
way, which is the heart of the general method for computing mass 
spectra in KK theories, and finally in Appendix C we list the tables 
of the various multiplets of the theory.

\section{$T^{11}$ geometry}

The $T^{pq}$ spaces \cite{Rom} are the coset manifolds $$ \frac{G}{H} 
= \frac{SU(2)\times SU(2)}{U_H(1)}, $$ where the $U_H(1)$ generator 
$T_H$ is embedded in the two $SU(2)$ factors as the
linear combination 
\beq 
\label{TH} 
T_H \equiv p \, T_3 + q \, \hat{T}_3 \qquad (p,q \in \hbox{{\msbm N},
coprime}) 
\eeq 
where $T_3$ and $\hat{T}_3$ generate $U(1)$ subgroups of the two 
$SU(2)$ in $G$.

To describe the geometry of these varieties, we can take two 
copies of the $SU(2)$ algebra with generators\footnote{
The conventions used for the $\e$ symbol, the group metrics etc. are 
reported in Appendix A.} $T_A$, $\hT_A$, 
($A=1\ldots3$):
\beq
\, [T_A,T_B] = \e_{AB}{}^C T_C, \qquad
\, [\hT_A,\hT_B] = \e_{AB}{}^C \hT_C,
\eeq
and write the Maurer--Cartan equations (MCe's) for their dual forms
$e^A$, $\hat{e}^A$:
\bea
d e^A + \dfrac{1}{2}\e_{BC}{}^A \, e^B e^C &=& 0, \\
d \hat{e}^A + \dfrac{1}{2}\e_{BC}{}^A \, \hat{e}^B \hat{e}^C &=& 0.
\eea
Choosing the two $U(1)$ subgroups of $SU(2)$ to be generated by
$e^3$ and $\hat{e}^3$, we may rewrite ($A=(i,3)$, $\hat{A} =
(r,3)$,$i,r=1,2$)
\begin{subequations}
\label{global1}
\bea
\label{MC1}
d e^i + {\e^i}_j \, e^3 e^j &=& 0, \\
\label{MC2}
d \hat{e}^r + {\e^r}_s \, \hat{e}^3 \hat{e}^s &=& 0,\\
\label{MC3}
d e^3 + \dfrac{1}{2} \e_{ij} \, e^i e^j &=& 0, \\
\label{MC4}
d \hat{e}^3 + \dfrac{1}{2}\e_{rs} \, \hat{e}^r \hat{e}^s &=& 0.
\eea
\end{subequations} 
Passing to the quotient $G \to G/H$, the MCe's for 
the left--invariant one--forms on $G$ become the corresponding MCe's 
for left--invariant forms on the coset $G/H$.  The linear combination
\beq
\omega \equiv \frac{p \, e^3 + q \, \hat e^3}{p^2 + q^2}
\eeq
dual to $T_H$ defined in \eqn{TH} can be identified as the 
$H$--connection of $G/H$, while $e^i$, $\hat e^r$ and the orthogonal 
combination
\beq
e^5 =  \frac{p \, e^3 - q \, \hat
e^3}{p^2 + q^2}
\eeq
can be identified with the five vielbeins
spanning the cotangent space to $G/H$.

We do not dwell with general values of $p$ and $q$ because 
we  
are interested only in $T^{pq}$ spaces endowed with supersymmetry and 
it has been shown in \cite{Rom} that this happens only if $p=q=1$.  
Thus we take 
\beq
T_{H} \equiv T_3 + \hT_3, \quad \omega \equiv \frac{e^3 + 
\hat{e}^3}{2},
\quad  e^5 \equiv \frac{e^3 - \hat{e}^3}{2},
\eeq
and
\beq
\label{T5}
T_5 = T_3 - \hat{T}_3,
\eeq
where $T_5$ is the coset generator dual to
$e^5$.
In the $(\omega,e^5)$ basis,  the MCe's \eqn{MC3}--\eqn{MC4} become
\bea
\label{omeMC}
d\omega + \frac{1}{4}
\e_{ij} e^i e^j -  \frac{1}{4} \e_{rs} \hat e^r \hat e^s &=& 0,\\
\label{e5MC}
d e^5 + \frac{1}{4} \e_{ij} e^i e^j +  \frac{1}{4} \e_{rs} \hat e ^r 
\hat e^s &=& 0.
\eea

It is convenient to introduce rescaled vielbeins
$V^a \equiv (V^i,V^{s},V^5)$:
\beq
\label{rescaled}
V^i = a \, e^i, \qquad
V^{s} = b\, \hat e^s,  \qquad
V^5 = c \, e^5,
\eeq
where $a,b,c$ are real rescaling factors which will be determined by
requiring that $T^{11}$ be an Einstein space \cite{libro,Cast}.
Using \eqn{rescaled} in \eqn{MC1}--\eqn{MC2} and 
\eqn{omeMC}--\eqn{e5MC}, we get
the full set of MCe's of $T^{11}$ for the vielbein $V^a$ and the
$H$--connection $\omega$
\bea
d V^i + \e^{ij} (\omega + c V^5) V_j &=& 0,\\
d V^r + \e^{rs} (\omega - c V^5) V_s &=& 0,\\
d V^5 + \frac{a^2}{4c} \e^{ij} V_i V_j - \frac{b^2}{4c} \e^{rs} V_r 
V_s &=& 0,\\
d \omega + \frac{a^2}{4} \e^{ij} V_i V_j + \frac{b^2}{4} \e^{rs} V_r 
V_s &=& 0.
\eea

Once we have the vielbeins, we may construct the Riemann connection
one--form $\cB^{ab} \equiv -\cB^{ba}$ ($a,b = i,s,5$), imposing the
torsion--free condition
\beq
\label{zerT}
d V^a - \cB^{ab} V_b = 0.
\eeq
It is straightforward to compare the equations one obtains
from \eqn{zerT}  with the rescaled MCe's, finding
\beq
\label{B}
\begin{array}{lcl}
\cB^{ij} = -\e^{ij} \left[ \omega + 
\left(c-\dfrac{a^2}{4c}\right)V^5\right], &&
\cB^{5i} = \dfrac{a^{2}}{4c} \, \e^{ij} \, V_j,  \\
\cB^{st} = -\e^{st} \left[ \omega - 
\left(c-\dfrac{b^2}{4c}\right)V^5\right], &&
\cB^{5s} = -\dfrac{b^{2}}{4c} \, \e^{st}\, V_{t},  \\
\cB^{is} = 0.&&
\end{array}
\eeq
The Riemann tensor  obtained from the
definition of the curvature 2--form:
\beq
R^{ab} \equiv d \cB^{ab} - \cB^a{}_c \cB^{cb} = {R^{ab}}_{cd} V^c V^d 
,
\eeq
has components given by
\beq
\label{Riemann}
\begin{array}{rcl}
{R^i}_{5j5} &=& - \delta^i_j \dfrac{a^2 b^2}{32 c^2}, 
\phantom{\dfrac{\stackrel{A}{A}}{\stackrel{A}{A}}} \\
{R^i}_{sjt} &=&  \dfrac{1}{8} \left(\dfrac{a^2 b^2}{4 c^2}\right) 
{\e^i}_j
\e_{st},  \phantom{\dfrac{\stackrel{A}{A}}{\stackrel{A}{A}}}\\
{R^i}_{jkl} &=& {\e^i}_j \e_{kl} \left( \dfrac{a^2}{2} - 
\dfrac{a^2b^2}{16 c^2}
\right)+ \dfrac{a^2b^2}{16 c^2} {\e^i}_{[k} \e_{l]j}, 
\phantom{\dfrac{\stackrel{A}{A}}{\stackrel{A}{A}}} \\
{R^s}_{5t5} &=& - \delta^s_t \dfrac{a^2 b^2}{32 c^2}, 
\phantom{\dfrac{\stackrel{A}{A}}{\stackrel{A}{A}}}\\
{R^s}_{trz} &=& {\e^s}_t \e_{rz} \left( \dfrac{a^2}{2} - 
\dfrac{a^2b^2}{16 c^2}
\right)+ \dfrac{a^2b^2}{16 c^2} {\e^s}_{[r} \e_{z]t}, 
\phantom{\dfrac{\stackrel{A}{A}}{\stackrel{A}{A}}} \\
{R^i}_{jst} &=&  \dfrac{a^2 b^2}{16 c^2} {\e^i}_j \e_{st}. 
\phantom{\dfrac{\stackrel{A}{A}}{\stackrel{A}{A}}}
\end{array}
\eeq
The requirement for $T^{11}$ to be an Einstein space means that its
Ricci tensor must be
\beq
\label{Ricci}
R^{a}{}_{b} = 2 \, e^2 \, \delta^a_b,
\eeq
where $e$ is the constant VEV of the IIB four--form self--dual field 
strength,
needed to obtain a spontaneous KK compactification via the 
Freund--Rubin mechanism \cite{FreundRubin}.  It is well known that 
$e^2$ is identified with the cosmological constant of $AdS_5$ (see 
\eqn{Romm} below).  The Ricci tensor 
components derived from \eqn{Riemann} are given by
\beq
{R^i}_{k} = \left(\frac{1}{2} a^2 - 
\frac{a^{4}}{16c^2}\right)\delta^i_k,
\qquad
{R^{s}}_{t} = \left(\frac{1}{2} b^2 -
\frac{b^{4}}{16c^2}\right)\delta^{s}_{t},
\qquad
{R^5}_{5} = \frac{a^4}{8c^2},
\eeq
and in order to satisfy \eqn{Ricci} we must have
\beq
\label{abc}
a^2 =b^2 = 6 e^2, \quad  \hbox{ and } \quad
c^2 = \frac{9}{4} e^2.
\eeq
This fixes the square of the rescalings and thus their 
absolute value, but not their sign.  We will see that this is going 
to be determined by supersymmetry requirement.

\bigskip
Before proceeding, we would like to briefly mention an alternative approach
for describing the geometry of the $T^{11}$ manifold  developed in \cite{Cast}
and used for example in \cite{G}.  If we decompose the $SU(2) \times
SU(2)$ Lie algebra {\msbm G} with respect to the $T_H$ generator as 
{\msbm G} = {\msbm H} + {\msbm K}, where {\msbm H} is made of the 
single $T_H$ generator and the coset algebra {\msbm K} contains the 
$T_i$, $\hat{T}_s$ and $T_5$ generators, the commutation relations 
between the {\msbm G} generators are 
\bea
\, [T_i,T_j] = \frac{1}{2} \e_{ij} (T_H + T_5), &&
\, [\hT_{s},\hT_{t}] = \frac{1}{2} \e_{st} (T_H - T_5), \nonumber \\
\label{Alg}
\, [T_5,T_i]=[T_H,T_i] = \e_i{}^j T_j, &&
\, [T_5,\hT_{s}]=[T_H,\hT_{s}] = \e_{s}{}^{t} \hT_{t}, \\
\, [T_i,\hT_{s}] &=& [T_5,T_H] = 0. \nonumber
\eea
From these, one can derive the structure constants
\bea
C_{ij}{}^H = C_{ij}{}^5  =\frac{1}{2} \e_{ij}, &&
C_{st}{}^H = C_{st}{}^5  = \frac{1}{2} \e_{st}, \nonumber \\
C_{5i}{}^j = C_{Hi}{}^j = \e_i{}^j, &&
C_{5s}{}^t = C_{Hs}{}^t = \e_{s}{}^{t}, \\
{C_{is}}^a &=& {C_{5H}}^a = 0. \nonumber
\eea
and find again the Riemann tensor components \eqn{Riemann} from the 
formula \cite{Cast}
\bea
{R^a}_{b,de} &=& \frac{1}{4} C^a_{bc} C^c_{de}
\left( \!\!\!
\begin{array}{c}\begin{array}{cc} a& b \end{array} \\ c \end{array}
\!\!\! \right) \frac{r(d) r(e)}{r(c)} +
\frac{1}{2} C^a_{bH} C^H_{de} r(d) r(e) \nonumber +\\
 &+& \frac{1}{8} C^a_{cd} C^c_{be}
\left(\!\!\!
\begin{array}{c}\begin{array}{cc} a& c \end{array} \\ d \end{array}
\!\!\!\right)
\left(\!\!\!
\begin{array}{c}\begin{array}{cc} b& c \end{array} \\ e \end{array}
\!\!\!\right) -
\frac{1}{8} C^a_{ce} C^c_{bd}
\left(\!\!\!\begin{array}{c}\begin{array}{cc} a& c \end{array} \\ e 
\end{array}
\!\!\!\right)
\left(\!\!\!\begin{array}{c}\begin{array}{cc} b& c \end{array} \\ d 
\end{array}
\!\!\!\right),
\eea
where
\beq
\left(\!\!\! \begin{array}{c}
\begin{array}{cc} a& b \end{array} \\ c \end{array}
\!\!\! \right) \equiv
\frac{r(a)r(c)}{r(b)} +\frac{r(b)r(c)}{r(a)}-\frac{r(a)r(b)}{r(c)},
\eeq
and $r(a) = \{ a,b,c\}$.

\section{Harmonic expansion}

\subsection{The general theory}

In this chapter we give a resum\'e of the general theory of harmonic 
expansion on coset manifolds (see e.g.  \cite{SS,Duffrev,libro,Sdd,Malt} and 
references therein)  adapt the general formulae  to 
our specific case, namely compactification of type IIB supergravity 
on the $T^{11}$ manifold.  

In Kaluza--Klein theories, we are faced with the problem of computing the mass
spectrum of a $D$--dimensional model compactified down to $D-d$ dimensions,
$d$ being the dimensions of a compact space, usually of the form of a coset
$G/H$.

We take the space--time  to be $AdS_{D-d}$, i.e. 
anti de Sitter space in $D-d$ dimensions.

The first thing to do is to 
compute the fluctuations of the $D$--dimensional fields around a 
particular background \cite{cdfpv,DaF}    which, in  spontaneous 
compactifications, turns out to be a solution of the $D$--dimensional 
theory equations of motion.  After the linearisation of the 
$D$--dimensional equations of motion has been performed, we are left, 
for each field of the theory, with an equation of the type: 
\beq
\label{duebox}
(\Box_x^{\{J\}} + \boxtimes_y^{[\l]}) \Phi^{\{J\}[\l]}(x,y) = 0.
\eeq
where $\Box$ and $\boxtimes$ are the kinetic operators in the $AdS$ 
space--time ($x$--coordinates) and in the compact space 
($y$--coordinates).  The label $\{J\}$ denotes the $AdS_{D-d}$ 
quantum 
numbers of the field (scalar, vector, spinor, etc.), while the 
$SO(d)$ 
representation in the tangent space to $G/H$ is labeled by the 
corresponding Young diagram\footnote{We will mostly omit the $\{J\}$ 
and $[\l]$ labels when not necessary.} $[\l] \equiv [\l_1, \ldots, 
\l_{[d/2]}]$.

The important thing about \eqn{duebox} is that the differential 
operator symbolically denoted by $\boxtimes_y^{[\l]}$ is a 
Laplace--Beltrami operator on $G/H$.  This means that, beside being 
invariant under $G$, its eigenfunctions, the harmonics on $G/H$, 
define a complete set of functions for the fields $\Phi^{[\l]}(y)$ on 
$G/H$ and, for each eigenvalue, they span an irreducible 
representation of $G$.

The invariant operators $\boxtimes$ can all be constructed in terms 
of the (flat) covariant derivatives $\cD_a$ ($a=1,\ldots d$) on $G/H$ 
and the invariant $G$--tensors.  Let us give a list of these 
operators 
for various irrepses $\left[\l_1, \ldots, \l_{[d/2]}\right]$ 
appearing in 
the KK compactification\footnote{Symmetrisation and 
antisymmetrisation 
are understood with weight one; we do not write explicitly the spinor 
index on the spinor harmonics: $\Xi \equiv \Xi^{\a}$.} 
\begin{subequations}
\label{iboxtimes}
\bea
\label{scalbox}
\boxtimes_y Y_{[0,0,\ldots,0]} &\equiv& \Box Y, \\
\label{1box}
\boxtimes_y Y_{[1,0,\ldots,0]} &\equiv& 2 \cD^a \cD_{[a} Y_{b]}, \\
\label{2box}
\boxtimes_y Y_{[1,\ldots,1,0,\ldots0]} &\equiv& (p+1) \cD^{a_{p+1}} 
\cD_{[a_1}Y_{a_2 \ldots a_{p+1}]}, \\
\label{2sym}
\boxtimes_y Y_{[2,0,\ldots,0]} &\equiv& 3 \cD^c \cD_{(c}Y_{ab)}, \\
\label{1/2op}
\boxtimes_y Y_{[1/2,1/2,\ldots,1/2]} &\equiv& \cD\!\!\!\!\slash \;
\Xi, \\
\label{3/2op}
\boxtimes_y Y_{[3/2,1/2,\ldots,1/2]} &\equiv& \gamma^{abc} \, \cD_b \;
\Xi_c.
\ees

Note that all the Laplace--Beltrami operators are either 
second--order 
differential operators acting on the bosonic harmonics $Y$ (which can 
be expressed in terms of the covariant Laplacian $\Box$ on $G/H$ plus 
curvature terms) or they are the first--order operators acting on 
spinor harmonics $\Xi$ (in this case they can be expressed in terms 
of the Dirac operator $\cD\!\!\!\!\slash$ on $G/H$).

We point out that for particular values of the internal space 
dimension $d=2k+1$, the Laplace--Beltrami operator acting on a 
$k$--form can be written in terms of the first--order operator $\star 
d$.  For example, in $d=5$, the Laplace--Beltrami operator on the 
2--forms $Y_{ab} V^a V^b$ is 
\beq
\label{firstord}
\boxtimes_y Y_{[1,1]} \equiv \star d Y_{ab} V^a V^b = \frac{1}{2} 
\e_{abcde} D^c Y^{de} \, V^a V^b,
\eeq
and the usual second--order operator is simply the square of the
first--order one.

\bigskip

To define the harmonics we start from the fundamental equation for
the coset representative $L(y)$ of $G/H$: \beq \label{fund} g L(y)
= L(y^{\prime})h(y,g) \eeq where $g \in G$, $h \in H$ and
$y,y^{\prime}$ are two points of $G/H$ related by the $g$ 
transformation.
In a given $G$--representation $\mD$ of indices $m$, $n$ we can 
rewrite
\eqn{fund} as follows
\beq
{\mD(g)^m}_n \mD^n(L(y))_{k_i} =
\mD^m(L(y^{\prime}))_{h_i}
 \mD(h(y,g))^{h_i}{}_{k_i}
\eeq
where the $N$--dimensional range of the indices $m$, $n$ of the
representation space of $\mD(h)$ has been fragmented into subsets 
$h_i$ corresponding to their $i$--th irreducible fragment; in other 
words, if $\{\nu\}$ identifies the $G$--representation and $\{\a_i\}$ 
identifies one of the irreducible $H$--representations according to 
the branching rule 
\beq
\{\nu\} \to \{\a_1\} \oplus \{\a_2\} \oplus \ldots \oplus
\{\a_M\},
\eeq
then we have
\beq
m = \{ h_1, \ldots , h_M\},   \qquad h_i =1,\ldots, n_i,\quad \quad 
\sum_1^M h_i = dim{\mD}.
\eeq
We  now define as irreducible harmonics
\beq
\left[Y^{\{\nu\}}_{\{\a_i\}}(y) \right]^m{}_{h_i} \equiv
{\mD^m}_{h_i}(L^{-1}(y)).
\eeq
The functions $Y^{(\nu)m}_{\{\a_i\}h_i}$, for fixed $\{\a_i\}$,
are a complete set of functions for the expansion of a field 
$\Phi_{h_i}(y)$
on $G/H$, where $\Phi_{h_i}(y)$ transforms in the irrep $\{\a_i\}$ of
$H$.
However, in KK, the $D$--dimensional physical fields also depend on 
the space--time coordinates $x$, so that a generic field 
$\Phi_{h_i}(x,y)$, transforming in the irrep $\{\a_i\}$, can be 
expanded as  
\beq
\label{laexp}
\Phi_{h_i}(x,y) = \sum_{\{\nu\}} \sum_m
\Phi^{\{\nu\}}_{m h_i}(x) (Y^{\{\nu\} }(y))^m_{h_i}.
\eeq
where for notational simplicity we have suppressed the index $\a_i$
referring to the particular $H$--representation.

Once the $G$--representation $\{\nu\}$ is fixed we can still have a 
state degeneration.
Indeed there are cases in which the same $G$--representation, with 
the same $H$ quantum numbers can be obtained in many ways.
In such cases the correct field expansion \eqn{laexp} is replaced by
\beq
\Phi_{h_i}(x,y) = \sum_{\{\nu\}} \sum_m \sum_{\delta}
\Phi^{\{\nu\} \delta}_{m h_i}(x) (Y^{\{\nu\} \delta}(y))^m_{h_i},
\eeq
where  $\delta$ counts the state degeneracy.

\bigskip

\begin{footnotesize}
We now exemplify the previous discussion with our specific
case, where the coset is
$G/H = (SU(2) \times SU(2))/U_H(1)$ and $dim(G/H) = 5$.
Thus $\{\nu\}$ is identified with $(j,l)$, the quantum 
numbers of the $SU(2)
\times SU(2)$ irreducible representations, and $\{\a_i\}$ is
identified with the charge $q_i$ of the $i$--th one--dimensional 
fragment
of the branching of a given $SU(2)\times SU(2)$ representation under 
$U_H(1)$.

To be more explicit, we write a generic representation of 
$SU(2) \times SU(2)$ in the Young tableaux formalism: 
\beq
\label{Tab1}
(j,l) \equiv \underbrace{\bet{|c|c|c|}\hline &$\ldots$ &
\\\hline\eet}_{2j}
\otimes \underbrace{\bet{|c|c|c|}\hline &$\ldots$ &
\\\hline\eet
}_{2l}.
\eeq
A particular component of \eqn{Tab1} can be written as
\beq
\label{Tab2}
 \underbrace{\bet{|c|c|c}\hline 1 &$\ldots$ & 1 \\\hline\eet}_{m_1}
\underbrace{\bet{|c|c|c|}\hline 2 &$\ldots$ & 2 \\\hline\eet}_{m_2}
\otimes
 \underbrace{\bet{|c|c|c}\hline 1 &$\ldots$ & 1 \\\hline\eet}_{n_1}
\underbrace{\bet{|c|c|c|}\hline 2 &$\ldots$ & 2 \\\hline\eet}_{n_2}
\eeq
and we have
\beq
\left\{ \bet{rcl} $2j$ &=& $m_1+m_2$ \\ $2j_3$ &=& $m_2-m_1$\eet 
\right. ,
\qquad
\left\{ \bet{rcl} $2l$&=& $n_1+n_2$ \\ $2l_3$ &=& $n_2-n_1$\eet 
\right. .
\eeq
Furthermore (recalling the definitions \eqn{TH}--\eqn{T5}) we get
\beq
\begin{array}{rcl}
T_H Y_{(q)}^{(j,l,r)} &=&{\rm i} \, q \, Y_{(q)}^{(j,l,r)}  \equiv 
{\rm i} \, (j_3 +l_3) \, Y_{(q)}^{(j,l,r)}, \\
T_5 Y_{(q)}^{(j,l,r)} &=& {\rm i} \, r \, Y_{(q)}^{(j,l,r)}  \equiv 
{\rm i}
\, (j_3 - l_3) \, Y_{(q)}^{(j,l,r)}.
\end{array}
\eeq
Hence
\bea
\label{qr}
\begin{array}{rcl}
2j_3 &=& q+r \equiv m_2-m_1, \\
2l_3 &=& q-r \equiv n_2-n_1.
\end{array}
\eea

It is easy to see that, given a generic $\{\nu\}=(j,l)$ 
$G$--representation and fixed the $H$ quantum number $q$, we have 
still the freedom to chose how to place the $1$ and $2$'s in the 
boxes of the Young tableaux.
We have therefore a state degeneration.
To remove such a degeneration we classify the various states by means 
of a fourth quantum number $r$ which is the charge of the state under 
the normalizer $U(1)$ in $H$.
This $U(1)$ coincides with the $R$--symmetry factor in the full 
isometry group $SU(2) \times SU(2) \times U(1)$.
We point out that this number is not a good quantum number for a 
fixed representation as it is not the same for all the $q$--fragments 
appearing in the expansion of a generic (spinor) tensor, but it is 
useful to determine the eigenvalues of the mass operators on the 
harmonics in terms of the known quantum numbers of the isometry group.

Since in this case the $h_i$ index is one--dimensional,
we can write the generic
harmonic of $T^{11}$ as $Y^{(j,l,r)}_{(q_i)}(y)^m$ and the expansion
of a field belonging to the subspace of charge $(q_i)$ as
\beq
\Phi_{(q_i)}(x,y) = \sum_{(j,l)} \sum_m \sum_{r}
\Phi^{(j,l,r)}_{q_i}(x)_m (Y^{(j,l,r)}(y))^m_{q_i}
\eeq
\end{footnotesize}

\bigskip

Coming back to the general case, we are now in position of 
constructing the
eigenfunctions of the Laplace--Beltrami operators acting on
the $SO(d)$ tangent group tensor and spinor fields.
Let us denote by $\Phi_{ab\ldots}^{[\l_1,\ldots,\l_{[d/2]}]}(x,y)$ 
such a field
where $[\l] \equiv [\l_1,\ldots,\l_{[d/2]}]$
are Young labels of the $SO(d)$
representation and $ab\ldots$ denote generically some tensor (or 
spinor) structure of the
indices.
We note that $H$ is a subgroup of $SO(d)$ (in our
example $H = U_H(1) \subset SO(5)$) and so we can
branch $[\l]$ with respect to $H$ obtaining a set of
$N$ irreducible representations of $H$
\beq
\label{embedding}
[\l] \stackrel{H}{\longrightarrow} \{\b_1\} + \{\b_2\} + \ldots
\{\b_N\}.
\eeq
To make this decomposition explicit we observe that if $G/H$ is a 
$d$--dimensional coset, then $H$ is a subgroup of $SO(d)$, its 
embedding being described by\footnote{ To simplify the notations we 
write $(T)^m{}_n$ instead of the appropriate notation for a 
representation $(\mD(T))^m{}_n$.} 
\beq
(T_H)^{m}{}_{n} = C_H{}^{ab} (T_{ab})^m{}_n
\qquad ({C_H}^{ab} = C_{Hc}{}^b \eta^{ca}),
\eeq
for a given $SO(d)$ representation labeled by indices $m$, $n$.
In general, any $SO(d)$ representation is fully reducible under $H$:
\beq
\label{emb2}
(T_H)^m{}_n = \left( \bet{cccc} $T_H^{m_1 n_1}$
& & & \\ &  $T_H^{m_2 n_2}$ & & \\
& & $\ddots$ & \\ & & &  $T_H^{m_M n_M}$\eet \right).
\eeq
In particular, the vielbein $V^a$, which is a $d$--dimensional $SO(d)$
vector ($[1,0,\ldots,0]$)
can be split into fragments transforming irreducibly under $H$ as 
follows:
\beq
\label{branchV}
V^a \to V^{h_1} \oplus \ldots \oplus  V^{h_N}.
\eeq

\bigskip

\begin{footnotesize}
Turning again to our case $G/H = T^{11}$, the set of generators of 
the 
isotropy group $H = U_H(1)$ is a single generator which is also named 
$T_H$.  Using the $SU(2) \times SU(2)$ algebra given in \eqn{Alg} we 
find the following decomposition of the $SO(5)$ vector and 
spinor representation with respect to the one--dimensional subgroup: 
for the vector representation we find 
\beq
\label{THv}
(T_H)_{ab} = C_{Hab} = \left(
\begin{array}{c|c|c}
\e_{ij} & &\\\hline
 & \e_{st}&\\\hline
 & &0
 \end{array}
\right),
\eeq
where each $\e^{ij} = \e^{st}= \left( \bet{cc} 0 & 1 \\ $-1$ & 0 
\eet\right)$,
while for the spinor representation
\beq
\label{THs}
(T_H) = C_H{}^{ab}(T_{ab})=-\frac{1}{4}
C_H{}^{ab}(\gamma_{ab}) = -\frac{1}{2}(\gamma_{12} +
\gamma_{34}) = i \left(
\bet{cccc} 0 &&& \\ &0&& \\ && 1 & \\ &&&-1 \eet
\right),
\eeq
where $\gamma_{ab}$ are the $SO(5)$ gamma matrices defined in the 
Appendix A.

Note that the vielbein $V^a$ breaks under $U_H(1)$ into five 
one--dimensional fragments $V^i = (V^1,V^2)$, $V^s = (V^3,V^4)$, 
$V^5$ 
with $U_H(1)$ charges given respectively by $(1,-1);(1,-1);0$.  
\end{footnotesize}

\bigskip

Each of the fragments appearing in \eqn{embedding} can be expanded in 
harmonics according to \eqn{laexp}.
Therefore the $SO(d)$ field $\Phi_{ab\ldots}^{[\l]}(x,y)$
(where $ab\ldots$ denote a set of indices labeling the $SO(d)$ 
representations
according to the Young symbol $[\l] = [\l_1,\ldots,\l_{[d/2]}]$)
can be expanded as 
\beq
\label{317}
\Phi_{ab\ldots}^{[\l]}(x,y) =
\left(
\bet{c} $\Phi_{h_1}(x,y)$ \\  $\Phi_{h_2}(x,y)$ \\ $\vdots$ \\  
$\Phi_{h_N}(x,y)$ \eet
\right)^{[\l]} =
 \sum_{\{\nu\}} \sum_m \sum_{\xi = 1}^N \Phi^{\{\nu\}}(x)_{m h_\xi}
\left(
\bet{c} $0$ \\  $\vdots$ \\  $Y_{h_\xi}(y)$ \\  $\vdots$ \\  $0$ \eet
\right)^{\{\nu\},m}
\eeq
where
\beq
\left(
\bet{c} $0$ \\  $\vdots$ \\  $Y_{h_\xi}$ \\  $\vdots$ \\  $0$ \eet
\right)
\eeq
is called a $SO(d)$ harmonic, the most general $SO(d)$ harmonic being
\beq
\label{3199}
Y_{ab\ldots}^{\{\nu\}m}(y) =
\left(
\bet{c} $Y_{h_1}$ \\  $Y_{h_2}$ \\ $\vdots$ \\  $Y_{h_N}$ \eet
\right)^{\{\nu\}m}.
\eeq

It is now evident that in the above expansion the set of irrepses 
$\{\nu\}$
of $G$ contributing to the expansion are only those that, when
branched with respect to $H$, contain at least one of the irrepses 
$\{\b_i\}$
appearing in the decomposition \eqn{embedding}.
In other words, for some $i$ and $j$ we must have $\{\a_i\} = 
\{\b_j\}$.

It may happen that $\{\nu\}$ contains the same $\{\a_\xi\}$ more
than once; in this case the index $\xi$ is extended to count also 
equivalent
copies of the same $\{\a_\xi\}$ contained in $\{\nu\}$.

\bigskip

At this point we can come back to the \eqn{duebox} equation.

Since $\boxtimes_y$ is an invariant Laplace--Beltrami operator on 
$G/H$
(invariant under the covariant Lie derivatives \cite{libro})
we can compute its action on the harmonics $Y_{i_\xi}^{\{\nu\}m}(y)$
obtaining
\beq
\boxtimes_y Y_{i_\xi}^{\{\nu\}m}(y) = M_{\xi\xi^{\prime}}^{\{\nu\}}
Y_{i_{\xi^{\prime}}}^{\{\nu\}m}(y)
\eeq
so that the linearised equations of motion of the $AdS$ fields become
\beq
(\delta_{\xi\xi^{\prime}} \Box_x + M_{\xi\xi^{\prime}}^{\{\nu\}})
\Phi^{\{\nu\}}_{n \xi^{\prime}}(x) = 0
\eeq
and by diagonalisation of the matrix $M_{\xi\xi^{\prime}}^{\{\nu\}}$ 
we find the eigenvalues for the various fragments
$\Phi^{\{\nu\}}_{n \xi^{\prime}}(x)$.

\subsection{The mass matrix}

Let us now discuss in some detail the computation of the mass matrix 
$M_{\xi\xi^{\prime}}^{\{\nu\}}$.  We want to show that the action of 
$\boxtimes_y$ as a differential operator on the harmonics can be 
reduced to a purely algebraic action in terms of generators of $G/H$ 
and $SO(d)$.

Since the Laplace--Beltrami operators are constructed in terms of the 
$SO(d)$ covariant derivatives 
\beq
\cD = d + \cB^{ab}T_{ab} \equiv d+\cB,
\eeq
where $T_{ab}$ are the $SO(d)$ generators,
setting $\cB = \omega^H + M$, ($\omega^H$ is defined in the
Appendix B) one can write
\beq
\label{deco}
\cD = \cD^H + M,
\eeq
where the H--covariant derivative is defined by
\beq
\label{DH}
\cD^H =  d + \omega^H.
\eeq

The usefulness of the decomposition \eqn{deco}--\eqn{DH},
lies in the fact that the
action of $\cD^H$ on the harmonics
can be computed algebraically.
Indeed one has quite generally (see Appendix B)
\beq
\label{DHH}
\cD^H = - r(a) T_a V^a.
\eeq

The covariant derivative, can then be written as 
\beq
\label{DL-1}
\cD  = (-r(a) V^a \, T_a + M^{ab}T_{ab}) .
\eeq

According to the branching of the $SO(d)$ fundamental representation
$[1,0,\ldots,0]$ of the vielbein \eqn{branchV},
we have in any $G$--representation
\beq
\cD_{h_i} = -r(i)  T_{h_i} + M^{ab}_{h_i} T_{ab},
\eeq
where $T_{ab}$ are the $SO(d)$ generators (whose normalization in the
vector representation is $(T_{ab})^{cd} = - \delta_{ab}^{cd}$ )
and $T_{h_i}$ are the generators of the coset algebra branched with 
respect to $H$.

\bigskip

\begin{footnotesize}
In our example we have $h_1 = i = \{1,2\}$, $h_2 = s= \{3,4\}$,
$h_3 = \{5\}$ and using  \eqn{zerT}--\eqn{B} and
\eqn{deco}--\eqn{DH}, the  $M$ matrices are
\bea
M^{ij} = -\left(c-\frac{a^2}{4c}\right) V^5 \e^{ij}, &&
M^{5i} = \frac{a^{2}}{4c} \e^{ij} V_j, \nonumber \\
\label{M}
M^{st} = \left(c-\frac{a^2}{4c}\right) V^5 \e^{st}, &&
M^{5s} = -\frac{a^{2}}{4c} \e^{st} V_{t}, \\
M^{is} &=& 0. \nonumber
\eea
and the covariant derivatives turn out to be
\bea
\cD_i &=& \left(-a T_i - \frac{a^2}{2c} \e_i{}^j T_{5j}
\right), \nonumber\\
\label{DaL-1}
\cD_{s} &=& \left(-a T_s + \frac{a^2}{2c} \e_{s}{}^{t}
T_{5t}\right),\\
\cD_5  &=& \left(-c T_5 - 2\left(c-\frac{a^2}{4c}\right)(T_{12} - 
T_{34})
 \right). \nonumber
\eea
\end{footnotesize}

\bigskip

Acting on a given harmonic $Y_{i_\xi}^{\{\nu\}m}(y)$, identified
by the irrep $\{\nu\}$ of $G$, with the covariant derivative, gives
\beq
\label{330}
\cD_{h_i} Y_{h_j}^{\{\nu\}m}  = - r(i) \left[ 
\mD(T_{h_i})\right]_{h_j h_k}
Y_{h_k}^{\{\nu\}m} + M_{a_i}^{ab} \left[  \mD(T_{h_i})\right]_{h_j 
h_k}
Y_{h_k}^{\{\nu\}m}.
\eeq
Note that the second term of the right hand side of \eqn{330} acts on 
the harmonics as an element of the $SO(d)$ Lie algebra, while the 
first term acts as an element of the Lie algebra of $G$, the isometry 
group.

Let us now observe that the action of $\cD_{h_i}$ on any harmonic is 
perfectly known once we define it on the basic harmonic $L^{-1}(y)$ 
in 
the fundamental representation of $G$ (if $G$ is an orthogonal group 
it is useful to consider the action of $\cD_{h_i}$ on the fundamental 
and spinor representations).  On the basis of this observation, the 
evaluation of the first term is usually done by identifying $\{\nu\}$ 
with a Young tableaux $[\l_1, \ldots, \l_{[d/2]}]$.  The action of 
$\mD(T_{h_i})$ on such a Tableaux is obtained by tensoring the action 
of the fundamental representation of $T_{h_i}$ on each box of the 
tableaux 
\beq
\begin{array}{|c|c|c|c|c|c|c|c|}\hline
i_1 & \ldots & \ldots & \ldots & \ldots & \ldots & \ldots & i_{\l_1} 
\\\hline
j_1 & \ldots  & \ldots & \ldots & \ldots & 
\multicolumn{1}{c|}{j_{\l_2}} \\\cline{1-6}
\vdots & \vdots & & \multicolumn{1}{c|}{\vdots} \\\cline{1-4}
 k_1 & \ldots & \multicolumn{1}{c|}{k_{\l_{[d/2]}}} \\\cline{1-3}
\end{array}
\eeq
and performing the required (anti-)symmetrisations.
Note that since in the fundamental representation
\beq
\mD(T_{h_i})_k{}^j \; \bet{|c|}\hline $j$ \\\hline \eet  = c_i \; 
\bet{|c|}\hline $k$
\\\hline \eet,
\eeq
$T_{h_i}$ can be thought as an operator destroying $\bet{|c|}\hline 
$j$ \\\hline \eet$ and creating
$\bet{|c|}\hline $k$ \\\hline \eet$ times a numerical coefficient 
(see the explicit examples of $SU(2)
\times SU(2)$ in the sequel).

The evaluation of the second term in \eqn{330} is simpler: it is 
sufficient to use the
matrix realization of $T_{ab}$ in the $[\l_1, \ldots, \l_{[d/2]}]$
representation of $SO(d)$.

Actually we are interested in the second order and first order 
Laplace--Beltrami operators given in \eqn{iboxtimes}.  The first 
order 
operators can be written in terms of 
\beq
\label{dslash}
\gamma^a \cD_a \equiv \sum_{i=1}^N \gamma^{a_i} \cD_{a_i}
=  \sum_{i=1}^N \gamma^{a_i} (-r(i) T_{a_i} + M^{ab}_{a_i}
T_{ab}),
\eeq
while the second order operators are all given, apart from curvature 
terms (which are known from the $G/H$ geometry), in terms of the 
covariant 
laplacian, namely:
\bea
\Box = \cD_{a} \cD^{a} &\equiv& \sum_i \cD_{a_i}  \cD^{a_i}
=  \sum_{i} (-r(i) T_{a_i} + M^{ab}_{a_i} T_{ab})
 (-r(i) T^{a_i} + M^{ab\ a_i} T_{ab}) = \nonumber \\
\label{boxx}
&=&  \sum_{i} (r(i)^2 T_{a_i}T^{a_i} -2 r(i) T_{a_i} M^{ab\ a_i} 
T_{ab})
 + M^{ab}_{a_i} M^{cd\ a_i} T_{ab} T_{cd}).
\eea
On an $SO(d)$ scalar harmonic which, being a $SO(d)$ singlet, is also 
an $H$--singlet, the previous expression assumes the remarkably 
simple 
form
\beq
\label{laplacian}
\Box Y_0^{\{\nu\}m} \equiv  \cD_{a} \cD^{a} Y_0^{\{\nu\}m}
=  \sum_{i} r(i)^2 T_{a_i}T^{a_i} Y_0^{\{\nu\}m},
\eeq
where $Y_0^{\{\nu\}m}$ is the one--dimensional $SO(d)$ {\it and} 
$H$--singlet.

\bigskip

\begin{footnotesize}
Formulae \eqn{dslash}--\eqn{laplacian} can be directly used in our 
example by setting $a_1 = \{1,2\} \equiv \{i\}$, $a_2 = \{3,4\} 
\equiv 
\{s\}$, $a_3 = \{5\}$, $r(i) = 6 \, e^2$, $r(s) = 6 \, e^2$, $r(5) = 
\frac{9}{2} \, e^2$, identifying the $T_{ab}$ with the $SO(5)$ 
generators and using the values of $M^{ab}_{a_i} = 
\{M^{ab}_{i},M^{ab}_{s},M^{ab}_{5}\}$ given in equations \eqn{M}.
\end{footnotesize}

\bigskip

Let us now show how to compute the spectrum of the physical masses 
for 
the $AdS$ fields appearing in the expansion \eqn{laexp}.  This is 
equivalent to compute the eigenvalues of the Laplace--Beltrami 
operators introduced before, the physical masses being the 
eigenvalues 
or simply related to them.

To compute the eigenvalues of a generic operator $\boxtimes$ we can 
compute either the explicit $dim [\l] \times dim [\l]$ numerical 
matrix obtained from the expression of the covariant Laplacian plus 
curvature terms obtained from \eqn{iboxtimes} after the action of the 
operators $T_{a_i}$, $T_{ab}$ in the representation $\{\a_i\}$ and 
$[\l]$ on the given harmonic has been evaluated; or, what happens to 
be easier in practice, one can think of \eqn{dslash}or \eqn{boxx} as 
a 
$dim [\l] \times dim [\l]$ matrix whose entries contain the operators 
$T_{a_i}$ and/or $T_{a_i}T^{a_i}$.  In that case we write: 
\beq
Y_{ab\ldots}^{\{\nu\}m} = \left(
\bet{c} $Y_{i_1}$ \\ $\vdots$ \\  $Y_{i_N}$ \eet \right)^{\{\nu\}m}
\eeq
and applying $\boxtimes_y$ to both sides of \eqn{laexp}
(suppressing for a moment the indices $\{\nu\} m$ since they remain 
inert in the computation)  we have
\beq
\boxtimes \Phi_{ab\ldots}^{[\l]} (x,y) =
\boxtimes \left\{ \sum_{\xi = 1}^N \Phi_{i_\xi}
 \left( \bet{c} $0$ \\ $\vdots$ \\  $Y_{i_\xi}$ \\ $\vdots$ \\$0$ 
\eet \right)
\right\} = \sum_{\xi = 1}^N \Phi_{i_\xi} \boxtimes_{\xi\xi^{\prime}}
 \left( \bet{c} $0$ \\ $\vdots$ \\  $Y_{i_{\xi^{\prime}}}$ \\
$\vdots$ \\$0$ \eet \right).
\eeq

Recalling that $\boxtimes_{\xi\xi^{\prime}}$ is a matrix of 
"operators" acting on the fragments $Y_{i_\xi}$ we have
\beq
\boxtimes_{\xi\xi^{\prime}} Y_{i_{\xi^{\prime}}} = M_{\xi\xi^{\prime}}
Y_{i_{\xi^{\prime}}}
\eeq
where $ M_{\xi\xi^{\prime}}$ is now a numerical entry.
Thus
\beq
\boxtimes Y_{ab\ldots}(x,y) = \left(
\bet{ccc}
$\Phi_{i_1} M_{11} Y_{i_1}$ & $\ldots$ & $\Phi_{i_1} M_{1N} Y_{i_N}$ 
\\
$\vdots$ & & $\vdots$ \\
$\Phi_{i_N} M_{N1} Y_{i_1}$ & $\ldots$ & $\Phi_{i_N} M_{NN} Y_{i_N}$
\eet
\right).
\eeq

By diagonalisation of $M_{\xi\xi^{\prime}}$ we finally find
\beq
\boxtimes \Phi_{ab\ldots}^{\{\nu\}} (x,y) =
\sum_{\{\nu\}m} \left( \bet{c}
$\l_1 \Phi_{i_1} Y_{i_1}$ \\
$\vdots$ \\
$\l_N \Phi_{i_N} Y_{i_N}$
\eet \right)^{\{\nu\}m} =
\sum_{\{\nu\}m} \sum_{\xi = 1}^N \Phi_{i_\xi}^{\{\nu\}m} \l_\xi
\left( \bet{c} $0$ \\ $\vdots$ \\  $Y^{\{\nu\}m}_{i_{\xi}}$ \\
$\vdots$ \\$0$ \eet \right)
\eeq
and we see that in general we have different eigenvalues
$\l_{\xi}$ in the different subspaces $\{\b_i\}$.

\subsection{Harmonic analysis on $T^{11}$}

Having explained the general procedure in the previous section, let 
us 
now restrict our attention to the harmonic analysis on $T^{11}$.

As we discussed in section 3.1, the first thing to do is to consider 
the background solution generating the $AdS_5 \times T^{11}$ geometry 
and compute the fluctuations of the fields around this solution.  
This 
latter is given by the following values of the ten--dimensional 
fields 
\cite{Rom}
\bea
F_{abcde}=  e \e_{abcde}, &&{R^a}_b = 2 e^2 \delta^a_{b}, \nonumber \\
\label{Romm}
F_{mnpqr}=  -e \e_{mnpqr}, &&{R^m}_n = -2 e^2 \delta^m_{n}, \\
B = A_{MN} = 0, &\qquad&\psi_{M}=\chi=0, \nonumber
\eea
where $F_{abcde}$ and $F_{mnpqr}$ are the projections on $T^{11}$ and 
$AdS_5$ of the ten--dimensional five--form $F$ defined as $F=dA_4$, 
$A_4$ being the real self--dual four--form of type IIB supergravity.  
The other fields of type $IIB$ supergravity are: the metric 
$G_{MN}(x,y)$ with internal and space--time components $g_{ab}(y)$, 
$g_{\mu\nu}(x)$ whose Ricci tensors in this background are given in 
\eqn{Romm} and the complex 0--form and 2--form $B$ and $A_{MN}$ (the 
fermionic fields $\psi_M$ and $\chi$ are obviously zero in the 
background \eqn{Romm}).

\bigskip

To determine the KK modes, one has first to determine the linearised 
field equations around the background \eqn{Romm} and then expand the 
excitations into the $T^{11}$ harmonics.  As explained in \cite{ 
cdfpv,KRV}, the general expansion can be simplified by choosing covariant 
gauge conditions in the internal space.

We are not going to give a detailed derivation of the linearised 
equations or gauge--fixing conditions for our case.  The right 
procedure has been explained in \cite{KRV} to which we refer.  There 
are however some subtleties and differences arising in making the 
same 
computations in our case which will be examined below.  To 
be more precise, while for most of the field excitations we can 
repeat 
the KK expansion and impose the gauge--fixing conditions as reported 
in \cite{KRV}, we have to be careful for the expansion of the modes 
deriving from the fourth--rank antisymmetric tensor $A_{MNPQ}$.

\bigskip

We name the ten--dimensional and five--dimensional fields as in
\cite{KRV} and we report in the Table 1 the relevant notations.
\begin{table}[ht]
\begin{center}
\vskip .3cm
 \begin{tabular}{|c|c|c|c|c|c|c|}\hline
 \hbox{Dim} & \multicolumn{5}{|c|}{\hbox{fields}}&
 \hbox{harm.}\\\hline
 \hline
10 D &$h^{\prime}_{\mu\nu}(x,y)$ & $h^a{}_a(x,y)$ &   
$A_{abcd}(x,y)$     &   $B(x,y)$  &   $A_{\mu\nu}(x,y)$&           \\
5  D &$H_{\mu\nu}(x)$ & $\pi(x)$             &   $b(x)$            
&   $B(x)$  &   $a_{\mu\nu}(x)$& $Y(y)$      \\\hline
10 D &$h_{a\mu}(x,y)$   & $A_{\mu abc}(x,y)$     &   $A_{\mu 
a}(x,y)$    &        &                &           \\
5  D &$B_\mu(x)$      & $\phi_\mu(x)$        &   $a_\mu(x)$        
&        &               & $Y_a(y)$     \\\hline
10 D &$A_{\mu\nu ab}(x,y)$&$A_{ab}(x,y)$         &                  
&        &               &           \\
5  D &$b^\pm_{\mu\nu}(x)$      &$a(x)$    &  & $ 
\phantom{\stackrel{A}{A}}$     &            & $Y_{[ab]}(y)$\\\hline
10 D &$h_{ab}(x,y)$     &                   &              &        
&          &            \\
5 D  &$\phi(x)$       &                &              &        
&               & $Y_{(ab)}(y)$\\\hline
\hline
10 D  &$\lambda(x,y)$    &$\psi_{(a)}(x,y)$           
&$\psi_\mu(x,y)$        &       &             &       \\
5 D   &$\lambda(x)$    &$\psi^{(L)}(x)$         &$\psi_\mu(x)$   &   
&   $ \phantom{\stackrel{A}{A}}$ &$\Xi(y)$    \\\hline
10 D  &$\psi_a(x,y)$      &                    &                 
&        &               &          \\
5 D  &$\psi^{(T)}(x)$  &     &          &        &       $ 
\phantom{\stackrel{A}{A}}$        &$\Xi_a(y)$  \\
\hline
 \end{tabular}
\caption{Fields appearing in the harmonic expansion.}
\end{center}
\label{fiel}
\end{table}

Using these notations the expansion of the fields, except $A_{MNPQ}$,
becomes
\begin{subequations}
\label{349}
\bea
h^{\prime}_{\mu\nu}(x,y) &=& \sum_{\{\nu\}} H_{\mu\nu}^{\{\nu\}} (x)
Y^{\{\nu\}}(y), \\
h_{\mu a}(x,y) &=& \sum_{\{\nu\}} B_{\mu}^{\{\nu\}} (x) 
Y_{a}^{\{\nu\}}(y), \\
h_{(ab)}(x,y) &=& \sum_{\{\nu\}} \phi^{\{\nu\}}(x) 
Y_{(ab)}^{\{\nu\}}(y), \\
{h^{a}}_{a}(x,y) &=& \sum_{\{\nu\}} \pi^{\{\nu\}}(x) Y^{\{\nu\}}(y), 
\\
A_{\mu\nu}(x,y) &=& \sum_{\{\nu\}} a_{\mu\nu}^{\{\nu\}} (x) 
Y^{\{\nu\}}(y), \\
A_{\mu a}(x,y) &=& \sum_{\{\nu\}} a_{\mu}^{\{\nu\}} (x) 
Y_{a}^{\{\nu\}}(y), \\
A_{ab}(x,y) &=& \sum_{\{\nu\}} a^{\{\nu\}} (x) Y_{[ab]}^{\{\nu\}}(y), 
\\
B(x,y) &=& \sum_{\{\nu\}} B^{\{\nu\}} (x) Y^{\{\nu\}}(y).
\ees
Note that in \eqn{349} do not appear derivative terms in the 
harmonics; this is due to the fact that using the gauge invariance 
freedom in the internal space, which we then fix by imposing 
\bes
\cD^{a} h_{(ab)} = \cD^{a} h_{a\mu} &=& 0, \\
\cD^{a} A_{ab} = \cD^{a} A_{a\mu} &=& 0,
\ees
we can restrict ourselves to the transverse harmonics, namely
harmonics satisfying the condition $\cD^a Y_{ab\ldots} = 0$.

In an analogous way, in the harmonic expansion of
the fourth--rank antisymmetric tensor fluctuations (denoted by 
$a_{MNPQ}$),
 we can choose the  covariant gauge conditions
\beq
\label{gaugecon}
\cD^{a}a_{abcd} = \cD^{a} a_{abc\mu} = \cD^{a} a_{ab\mu\nu} = \cD^{a} 
a_{a\mu\nu\rho} = 0,
\eeq
and remove again terms with gradients from the harmonics of the 
various
fluctuations.
This means that we can  expand these fluctuations in 
 KK modes ( we use the collective index
$\{\nu\}$ to denote the doublet $(j,l)$)
\begin{subequations}
\label{352}
\bea
a_{\mu\nu \rho \s}(x,y) = \sum_{\{\nu\}} 
b_{\mu\nu\rho\s}^{\{\nu\}}(x) Y^{\{\nu\}}(y), \\
a_{\mu \nu \rho a}(x,y) = \sum_{\{\nu\}} b_{\mu\nu\rho}^{\{\nu\}}(x) 
Y_a^{\{\nu\}}(y), \\
\label{amunuab}
a_{\mu\nu ab}(x,y) = \sum_{\{\nu\}} b_{\mu\nu}^{\{\nu\}}(x) 
Y_{ab}^{\{\nu\}}(y), \\
\label{amuabc}
a_{\mu abc}(x,y) = \sum_{\{\nu\}} \phi_{\mu}^{\{\nu\}}(x) 
Y_{abc}^{\{\nu\}}(y), \\
\label{aabcd}
a_{abcd}(x,y) = \sum_{\{\nu\}} b^{\{\nu\}}(x) Y_{abcd}^{\{\nu\}}(y).
\ees

We can achieve further simplifications if we consider the duality 
conditions in a five--dimensional space and the gauge conditions 
\eqn{gaugecon}.  From $\cD^{a} a_{abcd} = 0$, equation \eqn{aabcd} 
and 
since in five dimensions a 4--form is dual to a 1--form, 
we have 
\beq
\label{353}
\e^{abcde} \cD_d Y^{\{\nu\}}_e = 0,
\eeq
Now we recall that since the topology of $T^{11}$ is that of $S^2 
\times S^3$ its Betti numbers can be easily derived through the 
K\"unneth formula from the product of those of the spheres in the 
product. 
This tells us that there are no non--trivial one--cycles and so no closed 
non--exact 1--forms,
exactly as in the case of the $S^5$ compactification of \cite{KRV}.
Therefore \eqn{353} implies that $Y_e$ is an exact 1--form.
Thus we can simplify the $a_{abcd}$ expansion to
\beq
a_{abcd} = \sum_{\{\nu\}} b^{\{\nu\}} \e_{abcd}{}^{e} \, \cD_e 
Y^{\{\nu\}}.
\eeq

One could try to repeat the same procedure for the $a_{\mu abc}$ 
field, but here comes the point.
While this procedure is straightforward for the $S^{5}$ internal space
analyzed in \cite{KRV}, this is not the case for the $T^{11}$ 
manifold.
From the gauge condition $\cD^a a_{\mu abc} = 0$, equation 
\eqn{amuabc} and the duality relations,
we find
\beq
\label{e1}
\e^{abcde} \cD_c Y^{\{\nu\}}_{de} = 0,
\eeq
but now $Y_{de}$ can be a closed non--exact two--form since the
second and third Betti  numbers of $T^{11}$ are different
from zero and in fact
\beq
\label{Bettnum}
b_2 = b_3 = 1.
\eeq
This means that there exist a non--exact closed two--form on this 
space which can satisfy \eqn{e1}.

In \cite{Wittold} it has been shown that this single Betti 2--form
must be a $G$--singlet and thus there is only the 
$\{\nu\}=\{G\hbox{--singlet}\}=
\{0\}$ mode which
cannot be treated as in \cite{KRV}.
The $a_{\mu abc}$ expansion for $AdS_5 \times T^{11}$ is therefore
\beq
\label{e2}
a_{\mu abc} = \sum_{\{\nu\}} \left[ \phi_\mu^{\{\nu\}} \e_{abc}{}^{de}
\, \cD_d Y_e^{\{\nu\}} \right] + \tilde{\phi}_\mu^{(0)}
\e_{abc}{}^{de} Y_{de}^{(0)}
\eeq
which means that the only difference with the expansion in \cite{KRV} 
is the last term in \eqn{e2}.

Let us then analyze the consequences of this modification on the 
linearised
equations of motion.
It is easy to see that the only equations which are affected by this
change are the (E2), (M2) and (M3) equations of \cite{KRV}, namely
\bea
&&\frac{1}{2}(\Box_x + \Box_y) h_{\mu a} - \frac{1}{2} \cD_\mu 
\cD^{\rho}
h_{a\rho} - \frac{1}{2} \cD_{a} \cD^{\rho}h^{\prime}_{\mu\rho}
-\frac{4}{15} \cD_\mu \cD_{a} h^\gamma_\gamma
+ \nonumber \\
&+&\frac{1}{2}\cD_\mu \cD_{a} h^{\prime}{}^\s_\s - \frac{1}{2} \cD_\mu
\cD^{b} h_{(ab)} - \frac{1}{2} \cD_{a}\cD^{b}h_{\mu b} +
 = - \frac{e}{6} 
{\e_\mu}^{\nu\rho\s\tau}(\partial_{a}a_{\nu\rho\s\tau} -
4\partial_\nu a_{a\rho\s\tau}) + \nonumber \\
\label{E2}
&-& \frac{e}{6} {\e_{a}}^{bcde}
(\partial_\mu a_{bcde} - 4 \partial_{b}a_{\mu cde}) \\
\label{M2}
&& \partial_a a_{\mu\nu\rho\s} + 4 \partial_{[\mu}a_{\nu\rho\s]a} =
\frac{1}{4!} \e_{\mu\nu\rho\s a}{}^{\tau bcde}(\partial_\tau
a_{bcde} + 4\partial_{b}a_{cde\tau})+
e \e_{\mu\nu\rho\s}{}^\tau h_{a\tau} \\
\label{M3}
&& 3 \partial_{[\mu} a_{\nu\rho]ab}  + 2\partial_{[a}a_{b]\mu\nu\rho}=
\frac{10}{5!}\e_{\mu\nu\rho ab}{}^{\s\tau cde}
(3\partial_\gamma a_{de\s\tau} + 2 \partial_{\s} a_{\tau\gamma de})
\eea
The \eqn{E2} and \eqn{M3} equations now contain extra terms of the 
form
\beq
\e_a{}^{bcde} \partial_b a_{\mu cde} = \tilde{\phi}^0 \e_a{}^{bcde} 
\partial_b
\e_{cde}{}^{fg} Y^0_{fg} \sim \tilde{\phi}^0 \partial^b Y^0_{ab}
\eeq
which therefore reduce to zero due to the transversality condition on 
$Y_{ab}$
($\cD^a Y_{ab} = 0$) deriving from the gauge conditions 
\eqn{gaugecon} applied to \eqn{amunuab}.

We are left with the \eqn{M2} equation which, after insertion of the 
KK modes using \eqn{349}--\eqn{352}, contains in the r.h.s. a
new term in the following sector (equation (M2.2) of
\cite{KRV})
\beq
\label{e3}
3 \partial_{[\mu} b_{\nu\rho]}^0 - \frac{\Delta}{4} 
\e_{\mu\nu\rho}{}^{\s\tau} b^0_{\s\tau} =
2 \e_{\mu\nu\rho}{}^{\s\tau} \cD_\s \tilde{\phi}^0_\tau
\eeq
where $\Delta$ is the eigenvalue of the first--order operator $\star 
d$ on the two--form \eqn{firstord} and all the fields are in the 
singlet 
representation of the isometry group.
The divergence $\cD^\mu$ of this equation makes it independent on the 
new $\tilde{\phi}^0$
field due to the obvious identity $\cD_{[\mu} \cD_{\nu} 
\tilde{\phi}_{\rho]}^0 = 0$,
while we find a new equation of motion for the $\tilde{\phi}^0$ field
by contraction of \eqn{e3} with $\e^{\mu\nu\rho}{}_{\s\tau} \cD^{\s}$.
This yields
\beq
\label{260}
\cD^\s \cD_{[\s} \tilde{\phi}_{\tau]}^0 = \cD^\s b^0_{\s\tau}.
\eeq
\eqn{260} is easily seen to correspond to a new massless vector.
Indeed, since  $b^0_{mn}$ is
massive, i.e.
\beq
\e_{\mu\nu}{}^{\rho\s\tau} \cD_\rho b_{\s\tau}^0 = m_0 b_{\mu\nu}^0,
\eeq
its divergence vanishes $\cD^s b_{st} = 0$,
and \eqn{260} becomes
\beq
\cD^\s \cD_{[\s} \tilde{\phi}_{\tau]}^0 = 0.
\eeq

We see that the presence of non--trivial homology cycle on our 
manifold implies the presence of a massless vector in the singlet 
representation of the isometry group, and if supersymmetry is 
present, 
of an entire vector multiplet named Betti multiplet in \cite{DaF}.

\bigskip

Now that we have discussed the linearised equations of motion, we are 
in position to proceed to perform the harmonic expansion and thus to 
determine which irreducible representations $\{\nu\}$ of $SU(2) 
\times 
SU(2)$ do occur in \eqn{349}--\eqn{352}.

As explained in the general setting of section 2, in order to answer 
this question a preliminary step is to analyze the branching of the 
$SO(5)$ representations of the fields appearing in the l.h.s.  of 
\eqn{317} into $U_H(1)$ representations, according to the general 
formulae \eqn{embedding}--\eqn{emb2}

The internal $SO(5)$ representations we need to branch are: 
$$
\begin{array}{cll}
\ [0,0] & \hbox{ for the fields } h^{\prime}_{\mu\nu},h^a{}_a, 
A_{\mu\nu}, B &
 (\underline{1}) \\
\ [1,0] & \hbox{ for the fields } h_{\mu a}, A_{\mu a}, A_{\mu\nu\rho 
a},
A_{abcd} &
 (\underline{5}) \\
\ [1,1] & \hbox{ for the fields } A_{ab}, A_{\mu\nu ab}, A_{\mu abc} &
 (\underline{10}) \\
\ [2,0] & \hbox{ for the fields } h_{(ab)}&
 (\underline{14}) \\
\ [1/2,1/2] & \hbox{ for the fields } \l, \psi^\a_\mu &
 (\underline{4}) \\
\ [3/2,1/2] & \hbox{ for the fields } \psi^\a_a &
 (\underline{16})
\end{array}
$$
where on the extreme r.h.s. we have written the corresponding 
dimensions.

The explicit example worked out in \eqn{THv}, \eqn{THs} do already 
give the branching for the irrepses $\underline{5}$ and 
$\underline{4}$, namely 
\beq
\label{decco}
\begin{array}{rcll}
\hbox{\bf 5} &\to& 1 \oplus -1 \oplus 1 \oplus -1 \oplus 0 &
[\l_1,\l_2] = [1,0],\\
\hbox{\bf 4} &\to& 1 \oplus -1 \oplus 0 \oplus 0 \qquad  &
[\l_1,\l_2] = [1/2,1/2].
\end{array}
\eeq
where we have named the $U_H(1)$ irrepses by their charge.

From \eqn{decco} we easily find the analogous breaking law for
antisymmetric tensors ($[\l_1,\l_2] = [1,1]$), symmetric
traceless tensors ($[\l_1,\l_2] = [2,0]$) and spin tensors
($[\l_1,\l_2] = [3/2,1/2]$) by taking suitable combinations:
\beq
\label{360}
\begin{array}{rcll}
\hbox{\bf 10} &\to& \pm 1 \oplus \pm 1 \oplus \pm 2 \oplus 0 \oplus 0
\oplus 0 \oplus 0 &
[\l_1,\l_2] = [1,1], \\
\hbox{\bf 16} &\to& \pm 2 \oplus \pm 2 \oplus \pm 1 \oplus \pm 1
\oplus \pm 1 \oplus \pm 1 \oplus 0 \oplus 0 \oplus 0 \oplus 0 &
[\l_1,\l_2] = \left[\frac{3}{2},\frac{1}{2}\right], \\
\hbox{\bf 14} &\to& \pm 2 \oplus \pm 2 \oplus \pm 2 \oplus \pm 1 
\oplus \pm 1
 \oplus 0 \oplus 0 \oplus 0 \oplus 0 & [\l_1,\l_2] = [2,0].
\end{array}
\eeq

As it has been stressed in the sentence after \eqn{3199}, we know 
that 
the only irrepses $\{\nu\}$ of $SU(2) \times SU(2)$ which appear in 
the harmonic expansion of a field belonging to the $SO(d)$ irrep 
$[\l]$, are those which, when branched with respect to $U_H(1)$, 
contain at least one of the fragments of the branching \eqn{decco} or 
\eqn{360}.

To select such representations, we recall that a generic $G$ tableaux 
can be written as \eqn{Tab2}
$$
 \underbrace{\bet{|c|c|c}\hline 1 &$\ldots$ & 1 \\\hline\eet}_{m_1}
\underbrace{\bet{|c|c|c|}\hline 2 &$\ldots$ & 2 \\\hline\eet}_{m_2}
\otimes
 \underbrace{\bet{|c|c|c}\hline 1 &$\ldots$ & 1 \\\hline\eet}_{n_1}
\underbrace{\bet{|c|c|c|}\hline 2 &$\ldots$ & 2 \\\hline\eet}_{n_2}
$$
and that we have the \eqn{qr} relations
$$
\begin{array}{rcl}
q+r &\equiv& m_2-m_1, \\
q-r &\equiv& n_2-n_1.
\end{array}
$$
We observe that as long as $m_2-m_1$ and $n_2-n_1$ are even or 
odd, the same is true for $m_1+m_2$ and $n_1+n_2$.  Therefore the 
parity of $2j$ and $2l$ is the same as that of $2j_3$ and $2l_3$ and 
since $2j_3+2l_3= 2q$ can be even or odd, the same is true for 
$2j+2l$.  Thus $j$ and $l$ must either be both integers or 
both half--integers.  This means that the $q$ value of any $U_H(1)$ 
fragment of the $SO(5)$ fields is always contained in any 
$SO(5)$--harmonic in the irrep $(j,l)$ provided that $j$ and $l$ are 
both integers or half--integers.  Since $q+r$ and $q-r$ are related 
to 
the third component of the "angular momentum" of the two $SU(2)$ 
factors, one also has the conditions $|q+r| \leq 2j$ and $|q-r| \leq 
2l$.  The two above conditions select the harmonics appearing in the 
expansion.

\section{Computing the spectrum}

\medskip

{\it \large $\bullet$ The scalar harmonic}

\medskip

\noindent

The spectrum of the scalar harmonics 
$Y^{(j,l)}_{[0,0]}=Y_{q=0}^{(j,l,r)}$ is easily computed, since the 
relevant five--dimensional invariant operator is simply the covariant 
laplacian \eqn{scalbox}: 
\beq
\Box = \cD^a\cD_a \equiv  \cD^i\cD_i + \cD^{s}\cD_{s} + \cD^5\cD_5.
\eeq

Using \eqn{laplacian}
\beq
\label{box}
\Box Y_{q=0}^{(j,l,r)} = (-a^2 (T_i T_i + T_{s}T_{s}) - c^2 T_5
T_5) Y_{q=0}^{(j,l,r)},
\eeq

In order to evaluate \eqn{box}, we set
\bea
T_i = -\frac{i}{2} \s_i, &&
T_{s} = -\frac{i}{2} \hat{\s}_{s}, \\
T_5 =T_3 - \hT_3 &=& \frac{i}{2} (\hat{\s}_3 - \s_3), \nonumber
\eea
where $\s_A$ and $\hat{\s}_A$ are ordinary Pauli matrices.
Using the relations
\bea
\s_1 \bet{|c|}\hline 1 \\\hline\eet =  \bet{|c|}\hline 2 \\\hline\eet 
\qquad &
\s_2 \bet{|c|}\hline 1 \\\hline\eet =  -i \bet{|c|}\hline 2 
\\\hline\eet  &
\qquad \s_3 \bet{|c|}\hline 1 \\\hline\eet =  \bet{|c|}\hline 1 
\\\hline\eet \\
\s_1 \bet{|c|}\hline 2 \\\hline\eet =  \bet{|c|}\hline 1 \\\hline\eet 
\qquad &
\s_2 \bet{|c|}\hline 2 \\\hline\eet =  i \bet{|c|}\hline 1 
\\\hline\eet  &
\qquad \s_3 \bet{|c|}\hline 2 \\\hline\eet =  -\bet{|c|}\hline 2 
\\\hline \eet
\eea
(the same is true for $\hat{\s}$)
and observing that on a Young tableaux the $\s$'s act like a 
derivative
 (Leibnitz rule), we find on the first tableaux of \eqn{Tab2}
\bea
(\s_1\s_1 + \s_2\s_2)  \bet{|c|c|c|}\hline
 & \ldots & \\\hline
\eet &=& (2m_1(m_2+1) + 2m_2(m_1+1)) \bet{|c|c|c|}\hline
 & \ldots & \\\hline \eet = \\
&=& 4(j(j+1) - (j_3)^2)  \bet{|c|c|c|}\hline
 & \ldots & \\\hline \eet. \nonumber
\eea
An analogous result holds when acting with $\hat{\s}_1 \hat{\s}_1 + 
\hat{\s}_2
\hat{\s}_2$ on the second tableaux of \eqn{Tab2}, with 
$j\leftrightarrow l$.

The eigenvalue of $(\hat{\s}_3 - {\s}_3)^2$ on \eqn{Tab2} is
\beq
(m_2 - m_1 + n_2 -n_1)^2 = 4(j_3 + l_3)^2.
\eeq
For a scalar, $q=0$ and so, from \eqn{qr}, we have
\beq
j_3 = - l_3 = r/2,
\eeq
and we find
\beq
\Box Y_{(0)}^{(j,l,r)} = \left[a^2 j(j+1) +  b^2 l(l+1) +
(4c^2-a^2-b^2) \frac{r^2}{4}\right] Y_{(0)}^{(j,l,r)}.
\eeq
Substituting the values of $a$,$b$ and $c$ given  in \eqn{abc},
we obtain
\beq
\Box Y_{(0)}^{(j,l,r)} = H_0(j,l,r) Y_{(0)}^{(j,l,r)},
\eeq
where
\beq
\label{H0}
H_0(j,l,r)  \equiv 6\left(j(j+1) +  l(l+1) - \frac{r^2}{8}\right)
\eeq
is the eigenvalue of the Laplacian.
The same result was first given in \cite{G} using differential methods
and it was also obtained in \cite{JR}.

As shown in table \eqn{fiel}, the scalar harmonic appears in the 
expansion of the ten--dimensional fields $h^{\prime}_{\mu\nu}(x,y)$, 
$B(x,y)$, $h^{a}{}_{a}(x,y)$, $A_{abcd}(x,y)$ and $A_{\mu\nu}$.  The 
masses of the corresponding five--dimensional space--time fields are 
thus given in terms of $H_0(j,l,r)$, and explicitly they read
\bea
\label{acca}
m^2(H_{\mu\nu}) &=& H_0,\\
m^2(B) &=& H_0,\\
\label{pibi}
m^2(\pi,b) &=& H_0 + 16 \pm 8 \sqrt{H_0+4},\\
m^2(a_{\mu\nu}) &=&  8 + H_0 \pm 4 \sqrt{H_0+4}\ .\label{amunu}
\eea

Note that while the laplacian acts diagonally on the $AdS_5$ fields 
$H_{\mu\nu}(x)$ and $B(x)$, the eigenvalues for $\pi(x)$ and $b(x)$, 
which appear entangled in the linearised equations of motion 
 \cite{KW2}, \cite{KRV}, have been obtained after diagonalisation of a 
two by two matrix.

\bigskip

{\it \large $\bullet$ The spinor harmonic}

\bigskip

\noindent
We now give  the action of the $\cD\!\!\!\!\slash$ \ operator
\eqn{1/2op} on the spinor representation of $SO(5)$.
Equation \eqn{DaL-1} yields
\bea
\cD\!\!\!\!\slash &=& \gamma^a \cD_a = \gamma^i \left(-a T_i - 
\frac{a^2}{2c}
\e_{ij} T_5{}^j\right) +  \gamma^{s} \left(-a T_{s} + \frac{a^2}{2c}
\e_{st} T_5{}^{t}\right) +\nonumber \\
&+& \gamma^5 \left(-c T_5 -2 \left(c-\frac{a^2}{4c}
\right)(T_{12}-T_{34})\right),
\eea
where $T_{ab}$ are the $SO(5)$ generators in the spinor 
representation.
A straightforward computation gives
\beq
\label{matfer}
\cD\!\!\!\!\slash = \left(
\bet{cc}
$i c T_5 \unity_2 + \left(\dfrac{a^2}{4c} + c\right)\s^3$ & $-a 
\left(\s^i T_i
 + \s^3 \hT_1 - i \unity_2 \hT_2\right)$ \\
$a \left(\s^i T_i + \s^3 \hT_1 + i \unity_2 \hT_2\right)$ & $-ic T_5 
\unity_2$
\eet
\right).
\eeq

When substituting the values of $c$ and $a$ in the matrix 
\eqn{matfer} 
we note that \eqn{abc} defines them only up to a sign.  However, only 
one of them is consistent with supersymmetry.  Indeed, if a complex 
Killing spinor $\eta(y)$ generating $\cN = 2$ supersymmetry in 
$AdS_5$ 
is to exist, it must have the form
\beq
\label{eta}
\eta = \left(
\bet{c} $k$ \\ $l$ \\ 0 \\ 0
\eet
\right), \qquad k,l \in \hbox{\msbm C}
\eeq
since, being an $SU(2)\times SU(2)$ singlet, it must satisfy $T_H 
\eta = 0$
(see \eqn{THs}).
At this point the Killing equation
$\cD\!\!\!\!\slash \; \eta = \frac{5}{2} e \eta$ can be computed
from \eqn{matfer}  observing that on
an $SU(2) \times SU(2)$ singlet the $T_a$ generators have a null 
action and thus, using
\eqn{eta},
\beq
\cD\!\!\!\!\slash \; \eta = \left(
\bet{cc}
$\left(\dfrac{a^2}{4c} + c\right)\s^3$ & 0 \\
0 & 0
\eet
\right) \; \eta = \frac{5}{2} \, e \, \eta.
\eeq
This gives the correct value only if we choose $l=0$ and
\beq
c = - \frac{3}{2} e,
\eeq
while the sign of $a = \pm \sqrt{6} \, e$ is unessential.

Recalling the meaning of $c$ as the rescaling of the vielbein $V^5$, 
{\it we conclude that $T^{11}$ admits a Killing spinor, leading to 
$\cN = 2$ supersymmetry on $AdS_5$, only for one orientation of 
$T^{11}$}.

In order to compute the mass eigenvalues, we write \eqn{matfer} as an 
explicit $4 \times 4$ matrix whose entries are operators, according 
to 
the discussion given at the end of section 3.2
\beq
\label{espli}
\cD\!\!\!\!\slash = e \left(
\bet{cccc}
$-i\dfrac{3}{2} T_5 + \frac{5}{2}$ & 0 & $\sqrt{6} \hT_+$ & $\sqrt{6} 
T_-$ \\
0 & $-i\dfrac{3}{2} T_5 - \frac{5}{2}$  & $\sqrt{6} T_+$ & $-\sqrt{6} 
\hT_-$ \\
$-\sqrt{6} \hT_-$ & $-\sqrt{6} T_-$ & $\dfrac{3}{2}i T_5$ & $0$ \\
$-\sqrt{6} T_+$ & $\sqrt{6} \hT_+$ & 0 & $\dfrac{3}{2}i T_5$
\eet
\right),
\eeq
where we have set
\beq
T_\pm \equiv T_1 \pm i T_2, \qquad \hT_\pm \equiv \hT_1 \pm i
\hT_2.
\eeq
Note that \eqn{espli} acts on the four--dimensional spinor
representation of the $SO(5)$ spinor harmonic
$\Xi = \left(\bet{c}
$Y_{(0)}^{(j,l,r-1)}$ \\
$Y_{(0)}^{(j,l,r+1)}$ \\
$Y_{(-1)}^{(j,l,r)}$ \\
$Y_{(+1)}^{(j,l,r)}$
\eet\right)
$
which has been decomposed in one--dimensional $U_H(1)$ fragments
identified by their charge according to \eqn{THs}.  The operatorial 
matrix \eqn{espli} can now be replaced by a numerical one 
$M_{\xi\xi^{\prime}}$ according to the discussion of sect.  2, which 
is simply obtained from \eqn{espli} by substituting in each entry the 
value of the $T$--operators on the harmonics.  The fundamental 
substitutions one has to make are
\bea
T_+ Y_{(q)}^{(j,l,r)} &=& - i \left(j - \dfrac{q+r}{2}\right) 
Y_{(q+1)}^{(j,l,r+1)}\\
T_- Y_{(q)}^{(j,l,r)} &=& - i \left(j + \dfrac{q+r}{2}\right) 
Y_{(q-1)}^{(j,l,r-1)}\\
\hat T_+ Y_{(q)}^{(j,l,r)} &=& - i \left(l - \dfrac{q-r}{2}\right) 
Y_{(q+1)}^{(j,l,r-1)}\\
\hat T_- Y_{(q)}^{(j,l,r)} &=& - i \left(l + \dfrac{q-r}{2}\right)  
Y_{(q-1)}^{(j,l,r+1)}\\
T_5 Y_{(q)}^{(j,l,r)} &=& i r  Y_{(q)}^{(j,l,r)}
\eea
From this action we obtain
\beq
\left(
\begin{array}{cccc}
1 +  \dfrac{3}{2}r & 0 & -i \sqrt{6} \left(l + \dfrac{r}{2} + 
\dfrac{1}{2}\right) &
   -i \sqrt{6}\left(j + \dfrac{r}{2} + \dfrac{1}{2}\right) \\
    0 &   -1 + \dfrac{3}{2}r & -i \sqrt{6}\left(j -
   \dfrac{r}{2} + \dfrac{1}{2}\right)&
   i \sqrt{6} \left(l- \dfrac{r}{2} + \dfrac{1}{2}\right) \\
   i \sqrt{6} \left(l- \dfrac{r}{2} + \dfrac{1}{2}\right)&i 
\sqrt{6}\left(j + \dfrac{r}{2} + \dfrac{1}{2}\right) &
   -\dfrac{3}{2}r& 0\\
 i \sqrt{6}\left(j -   \dfrac{r}{2} + \dfrac{1}{2}\right)&-i \sqrt{6} 
\left(l + \dfrac{r}{2} + \dfrac{1}{2}\right)& 0&
  - \dfrac{3}{2}r
\end{array}
\right)
\eeq

Diagonalising now this matrix, we get the  eigenvalues
\beq
\label{autof}
\lambda_{[\frac{1}{2},\frac{1}{2}]}=\left\{
 \left(\frac{1}{2} \pm \sqrt{H_0(r-1)+4}\right) ,
 \left(- \frac{1}{2} \pm \sqrt{H_0(r+1)+4}\right)\right\},
\eeq
where by $H_0(r\pm 1)$ we mean $H_0(j,l,r\pm 1)$.
To \eqn{autof} we have to add the four eigenvalues  
obtained from the inequivalent mass matrix, where one replaces $r$ by 
$-r$.

The masses for the spinors and gravitinos are given in terms of
$\cD\!\!\!\!\slash$ by a numerical shift
\beq
\label{camp}
\begin{array}{rrcl}
\hbox{ gravitino : } & m(\psi_\mu ) &=& \cD\!\!\!\!\slash -
\dfrac{5}{2}; \\
\hbox{dilatino : } & m(\l) &=& \cD\!\!\!\!\slash +1; \\
\hbox{longitudinal spinors: } & m(\psi^{(L)}) &=&
 \cD\!\!\!\!\slash +3.
\end{array}
\eeq

This part of the spectrum has been first calculated in \cite{JR} and 
the values for the masses of these states agree with the ones we have 
written above.

\bigskip

{\it \large $\bullet$ The vector harmonic}

\bigskip

\noindent
The Laplace--Beltrami operator on the vector harmonics \eqn{1box} is
\beq
\label{vecop}
\boxtimes Y_a = \Box Y_a + 2 {R_a}^b Y_b.
\eeq

As can be easily seen from \eqn{THv} to decompose a vector index
under $U_H(1)$, it is convenient to go to a complex basis, defining 
$(\pm) = 1 \pm i 2$ and $(\hat{\pm}) = 3 \pm i 4$.

In this new basis, the operator \eqn{vecop} becomes the 
matrix ($e = 1$) 
$$
\left(
\bet{ccccc}
$\Box + \frac{21}{4}  + \frac{3}{2}i  T_5$ & & & & $\sqrt{6} i T_+$ \\
& $\Box + \frac{21}{4}  - \frac{3}{2}i  T_5$  & & & $-\sqrt{6} i T_-$ 
\\
&& $\Box + \frac{21}{4}  - \frac{3}{2}i  T_5$  & & $-\sqrt{6} i
\hat T_{ +}$ \\
&&& $ \Box + \frac{21}{4}  + \frac{3}{2}i  T_5$  & $\sqrt{6} i
\hat T_{-}$ \\
$\frac{\sqrt{6}}{2} i T_-$ & $-\frac{\sqrt{6}}{2} i T_+$ &
$-\frac{\sqrt{6}}{2} i \hat T_{ -}$ &$\frac{\sqrt{6}}{2} i \hat 
T_{+}$ &
$\Box + 8 $
\eet
\right)
$$
acting on the  harmonics
\beq
\label{vechar}
Y_a^{\{\nu\}} = \left(\bet{c}
$Y_{(+)}$ \\
$Y_{(-)}$ \\
$Y_{(\hat +)}$ \\
$Y_{(\hat -)}$\\
$Y_{(0)}$
\eet\right) =
\left(\bet{c}
$Y_{+1}^{(j,l,r+1)}$ \\
$Y_{-1}^{(j,l,r-1)}$ \\
$Y_{+1}^{(j,l,r-1)}$ \\
$Y_{-1}^{(j,l,r+1)}$\\
$Y_{0}^{(j,l,r)}$
\eet\right).
\eeq

Note that, in principle, the five entries of \eqn{vechar} are not 
independent because of the transversality condition $\cD^a Y_a = 0$.  
However, it turns out to be more convenient to treat the five 
harmonics as independent, which amounts to say that we now consider 
also longitudinal harmonics of the form $\cD_a Y$.  The presence of a 
longitudinal harmonic means that among the five eigenvalues of the 
matrix, we should find the eigenvalue of the laplacian on the scalar 
harmonic, since the Laplace--Beltrami on a longitudinal $p$--form 
harmonic has the same eigenvalues of the Laplace--Beltrami operator 
acting on the $(p-1)$--form harmonic.  In our case, we should then 
find the eigenvalue of the laplacian $H_0$ between the five we get.
This indeed is the case and the remaining four eigenvalues for the 
(transverse) 1--forms are
\beq
\lambda_{[1,0]}=\{3 + H_0(j,l,r\pm 2),  H_0 + 4 \pm 2
\sqrt{H_0+4}\}.
\eeq
The mass spectrum of the sixteen vectors is thus
\bea
\label{amu}
m^2(a_{\mu}) &=& \left\{\begin{array}{c}
3 + H_0(j,l,r\pm 2) \\
 H_0 + 4 \pm 2 \sqrt{H_0+4} \end{array} \right. ,\\\label{bifi}
m^2(B_{\mu},\varphi_{\mu}) &=& \left\{\begin{array}{l}
H_0(j,l,r\pm 2) +7 \pm 4 \sqrt{H_0+4} \\
 H_0 + 12 \pm 6 \sqrt{H_0+4} \\
 H_0 + 4 \pm 2 \sqrt{H_0+4}
\end{array} \right.
\eea

Actually, as the Laplace--Beltrami operator acts diagonally on the 
complex vector field $a_{\mu}(x)$, so we get eight mass values .  
Conversely, the vectors $B_{\mu}(x)$, $\varphi_{\mu}(x)$ get mixed in 
the linearised equations of motion \cite{KRV}, and upon 
diagonalisation we find two extra masses for each eigenvalue.  As for 
the scalar fields, we will use the same names $B_\mu$ and $\phi_\mu$ 
also for the eigenstates corresponding to linear combinations with 
plus or minus sign respectively in the mass formulae \eqn{bifi} .

\bigskip

{\it \large $\bullet$ The two--form harmonic}

\bigskip

\noindent
The relevant Laplace--Beltrami operator is now of the first--order
\eqn{firstord} and can be simply expressed as
\beq
\label{dueop}
\frac{1}{2}\e^{abcde} \cD_c Y_{de} = \frac{1}{2} \e^{abcde} T_c 
Y_{de} +  \e^{abcde} (M_c)_d{}^s
Y_{se}.
\eeq

For the purpose of the computation it is useful to think the
action of the generators in the representation space of a vector
$Y_{ab}$ labeled by a couple of antisymmetric indices.

Again, it is simpler to use a complex basis, where the various
components of the tensor have a definite $U_H(1)$ charge.
Ordering the ten components of $Y_{ab}$ as follows
\beq
\label{vecdue}
\left(
\bet{c}
$Y_{+\hat{+}}$ \\
$Y_{5+}$ \\
$Y_{5\hat{+}}$ \\
$Y_{+-}$ \\
$Y_{\hat{+}\hat{-}}$ \\
$Y_{+\hat{-}}$ \\
$Y_{-\hat{+}}$ \\
$Y_{5-}$ \\
$Y_{5\hat{-}}$ \\
$Y_{-\hat{-}}$
\eet
\right) =
\left(
\bet{c}
$Y_{+2}^{(j,l,r)}$ \\
$Y_{+1}^{(j,l,r+1)}$ \\
$Y_{+1}^{(j,l,r-1)}$ \\
$Y_{0}^{(j,l,r)}$ \\
$Y_{0}^{(j,l,r)}$ \\
$Y_{0}^{(j,l,r+2)}$ \\
$Y_{0}^{(j,l,r-2)}$ \\
$Y_{-1}^{(j,l,r-1)}$ \\
$Y_{-1}^{(j,l,r+1)}$ \\
$Y_{-2}^{(j,l,r)}$
\eet
\right)
\eeq
and decomposing the free indices $ab$ of \eqn{dueop} as in
\eqn{vecdue}, the operator gives rise to the 
ten--dimensional vector
\beq
\label{azione2}
\left(
\bet{c}
$-2 \sqrt{3} T_{\hat +} Y_{5+} + 2 \sqrt{3} T_{+} Y_{5\hat+} +
\frac{3}{\sqrt{2}} T_5 Y_{+\hat +}$ \\
$2 i \sqrt{2} Y_{5+} + \sqrt{3} T_{\hat +} Y_{+\hat -} -  \sqrt{3} 
T_{\hat -} Y_{+\hat +} - \sqrt{3} T_+ Y_{\hat{+}\hat{-}}$ \\
$-2 i \sqrt{2} Y_{5\hat+} - \sqrt{3} T_{+} Y_{-\hat +} -  \sqrt{3} 
T_{\hat +} Y_{+-} + \sqrt{3} T_- Y_{+\hat{+}}$ \\
$2 \sqrt{3} T_{\hat +} Y_{5 \hat -} - 2 \sqrt{3} T_{\hat -} Y_{5 \hat 
+} + \frac{3}{\sqrt{2}} T_5 Y_{\hat{+}\hat{-}}$ \\
$2 \sqrt{3} T_{+} Y_{5-} - 2 \sqrt{3} T_{-} Y_{5+} + 
\frac{3}{\sqrt{2}} T_5 Y_{+-}$ \\
$-2 \sqrt{3} T_{+} Y_{5 \hat -} + 2 \sqrt{3} T_{\hat -} Y_{5+} - 
\frac{3}{\sqrt{2}} T_5 Y_{+\hat{-}} +3 \sqrt{2} i Y_{+\hat{-}}$ \\
$2 \sqrt{3} T_{\hat +} Y_{5-} - 2 \sqrt{3} T_{-} Y_{5\hat +} - 
\frac{3}{\sqrt{2}} T_5 Y_{-\hat{+}} -3 \sqrt{2} i Y_{-\hat{+}}$ \\
$-2i \sqrt{2} Y_{5-} - \sqrt{3} T_{\hat +} Y_{-\hat -} +  \sqrt{3} 
T_{\hat -} Y_{-\hat +} + \sqrt{3} T_- Y_{\hat{+}\hat{-}}$ \\
$2i \sqrt{2} Y_{5\hat -} + \sqrt{3} T_{+} Y_{-\hat -} +  \sqrt{3} 
T_{\hat -} Y_{+-} - \sqrt{3} T_- Y_{+\hat{-}}$
\\
$-2 \sqrt{3} T_{\hat -} Y_{5-} + 2 \sqrt{3} T_{-} Y_{5\hat-} +
\frac{3}{\sqrt{2}} T_5 Y_{-\hat -}$
\eet
\right).
\eeq

Comparing \eqn{azione2} with \eqn{vecdue}, one reconstructs the
operator matrix $\boxtimes_{\xi\xi^{\prime}}$ and then one
computes the numerical matrix $M_{\xi\xi^{\prime}}$.  From this 
latter, we get four 0 eigenvalues corresponding to the longitudinal 
harmonics\footnote{Indeed in this case on a longitudinal harmonic 
$\cD_{[a} Y_{b]}$ the operator is identically zero, i.e. $\e^{abcde} 
\cD_c 
\cD_d Y_e = 0$.} and six eigenvalues corresponding to the transverse 
ones 
\beq
\lambda_{[1,1]}=\left\{i\left(1\pm \sqrt{H_0(j,l,r\pm 2) + 4}\right),
\pm i \sqrt{H_0 +4}\right\}.
\eeq
The corresponding masses for the physical states are
\bea
m^2(b_{\mu\nu})& =&\left\{
\begin{array}{l}
H_0+4\\
H_0+4\\
5+H_0(j,l,r\pm2)\pm 2\sqrt{H_0(j,l,r\pm2)+4}
\end{array}
\right. ,\\
m^2(a) &=& \left\{\begin{array}{l}
H_0 +4 \pm 2 \sqrt{H_0+4} \\
 H_0(j,l,r\pm 2) + 1 \pm 2 \sqrt{H_0(j,l,r\pm 2)+4}
\end{array} \right. .
\eea

Also in this case part of the spectrum we have shown has been
computed in \cite{JR} and it agrees with our results.

\bigskip

{\it \large $\bullet$ The other harmonics}

\bigskip

\noindent
We have not calculated either the eigenvalues of 
$\cD\!\!\!\!\slash$
corresponding to the vector--spinor harmonic $\Xi_a$ which produce 
$AdS_5$ spinors
$\psi^{(T)}$, or the eigenvalues of the symmetric traceless harmonic
$Y_{(ab)}^{(\nu)}$.
However, we know a priori how many states we obtain in these two 
cases, and by a counting argument we can circumvent the problem of 
the 
explicit computation of the eigenvalues of their mass matrices.
For the vector--spinors we have in principle a matrix of rank 20, 
that 
becomes $16\times 16$ due to the irreducibility condition, and 
further 
gets to $12\times 12$, once the transversality condition $\cD^a \Xi_a 
= 0$ is imposed.
In this way we are left with $12$ non--trivial (non longitudinal) 
eigenvalues and thus we expect $12$ $\psi^{(T)}$ spinors.  In an 
analogous way, the traceless symmetric tensor $Y^{(\nu)}_{(ab)}$ 
gives 
a $14\times 14$ mass--matrix out of which five eigenvalues are 
longitudinal leaving $9$ non--trivial eigenvalues .

If we match the bosonic and fermionic degrees of freedom including 
the 
$12 + 12$ (right) left--handed spinors $\psi^{(T)}$ and the $9$ real 
fields $\phi$ of the traceless symmetric tensor we find $128$ bosonic 
degrees of freedom and $128$ fermionic ones.
Therefore, once we have correctly and unambiguously assigned all the 
fields except the $\psi^{(T)}$ and $\phi$ to supermultiplets of 
$SU(2,2|1)$, the remaining degrees of freedom of $\psi^{(T)}$ and 
$\phi$ are uniquely assigned to the supermultiplets for their 
completion.

\section{Filling of $SU(2,2|1)$ multiplets}

This section aims at combining the results of the harmonic expansion 
on $T^{11}$ with those of the purely group theoretical analysis of 
the 
$SU(2,2|1)$ representations, and proceed by filling $SU(2,2|1)$ 
supermultiplets with the appropriate eigenvalues of the KK mass 
operators.  This procedure was originally devised \cite{CFN} in the 
analysis of the spectra of $\cN=2$
supersymmetric $AdS_4$ compactifications of eleven dimensional 
supergravity , where the full symmetry group is $OSp(4|2)$, and has 
recently been revisited in
\cite{M111}, leading to the uncovery of an interesting structure of 
new short multiplets.  In the $AdS_4$ case, the reconstruction of the 
supermultiplets was advantaged by the results of \cite{DaF} where 
universal mass relations among fields bound to each other by 
supersymmetry transformations were derived from the general 
properties 
of harmonic expansion on coset manifolds with Killing spinors 
\cite{pfre}.  For $AdS_5$ compactifications this tool is not 
available, but, as we will see, one can still fully assemble all the 
KK 
fields into multiplets and retrieve elsewhere all the necessary 
information.

On the group theory side, we need the unitary highest weight 
representations of $SU(2,2|1)$, originally worked out in \cite{FF1}, 
\cite{DP} and recently nicely popularized in an appendix of 
\cite{FGPW}.  They are characterized in terms of the quantum numbers 
of the $U(1)\times SU(2)\times SU(2)\subset SU(2,2)\subset SU(2,2|1)$ 
bosonic subalgebra $(E_0,s_1,s_2)$, where $E_0$ is the $AdS$ energy, 
and by the internal symmetry $SU(2)\times SU(2)\times U(1)$ labels 
$(j,l,r)$, $r$ being the $R$-symmetry charge.

The general relations between $E_0$ and the masses for fields of 
various spin are
\bea
\hbox{spin 2:} & E_0^{(2)} = & 2 + \sqrt{4+m_{(2)}^2} \nonumber \\
\hbox{spin 3/2:} & E_0^{(3/2)} = & 2 + |m_{(3/2)}+3/2| \nonumber \\
\label{Delta}
\hbox{spin 1:} & E_0^{(1)} = & 2 + \sqrt{1+m_{(1)}^2}  \\
\hbox{two--form:} & E_0^{(2f)} = & 2 + |m_{(2f)}| \nonumber \\
\hbox{spin 1/2:} & E_{0}^{(1/2)} = & 2 \pm |m_{(1/2)}| \nonumber \\
\hbox{spin 0:} & E_{0}^{(0)} = & 2 \pm \sqrt{4+m_{(0)}^2} \nonumber
\eea
The sign ambiguity in the spin zero and
1/2 formulae occurs because in these specific cases
 the unitarity bound $E_0 \geq 1 +s$ allows the
possibility $E_0 < 2$.

We will arrange our results in a serie of nine tables , summarizing 
the properties of the various families of unitary irreducible 
$SU(2,2|1)$
representations ${\cal D}(E_0,s_1,s_2;r)$, each generated by a 
specific
$(s_1,s_2)$ state, having $E_0^{(s)}=E_0$ and $R$--charge $r$. All 
descendant
states have an $E_0^{(s)}$ value shifted in a range of $\pm 2$ (in 
1/2 steps) with respect to the $E_0$ of the multiplet, while their 
$R$--symmetry is shifted in a range of $\pm 2$ in integer steps.
The highest spin state has unshifted $R$--charge $r$.  We have one 
graviton multiplet ${\cal D}(E_0,\frac{1}{2},\frac{1}{2};r)$(table 
2), 
two left ${\cal D}(E_0,\frac{1}{2},0;r)$ and two right ${\cal 
D}(E_0,0,\frac{1}{2};r)$ gravitino multiplets (tables 3-6) and four 
vector multiplets ${\cal D}(E_0,0,0;r)$ (tables 7-10).
Thus their structure and the information in the columns labeled 
$(s_1,s_2)$, $E_0^{(s)}$ and $R$-symm.  is purely group theoretical 
and implied by the analysis of \cite{FF1,DP,FGPW}.

For generic values of the $SU(2)\times SU(2)$ quantum numbers $j,l$ 
and the $R$--symmetry $r$, the multiplets of the Tables 2--10 are 
massive long multiplets of $SU(2,2|1)$.  However, it is well known 
\cite{FGPW} that multiplet shortening occurs for specific values of 
the $SU(2,2|1)$ quantum numbers, when $s_1s_2=0$ or unitarity bounds 
are saturated.
For $SU(2,2|1)$, there are three types of shortened representations: 
\begin{itemize}
\item[$\diamond$]  massless $AdS$ multiplets,
\beq
\label{52}
E_0=2+s_1+s_2, \qquad\qquad (s_1-s_2)=\frac{3}{2}r,
\eeq

\item[$\star$]   semi--long $AdS$ multiplets:
\beq
\label{53}
E_0=
\left\{\begin{array}{c}
\dfrac{3}{2} r + 2 s_2+2 \\
-\dfrac{3}{2} r+2 s_1+2
       \end{array} \right.
\eeq
\item[$\bullet$]  chiral $AdS$ multiplets
\beq
\label{54}
E_0=\left|\frac{3}{2} r\right|.
\eeq
\end{itemize}
These shortened representations are the most interesting ones in 
light 
of the correspondence with the CFT at the boundary, and their field 
theoretical counterparts have been extensively discussed in 
\cite{CDD}.  
Shortening has been related to peculiar $E_0$ values reported in 
\eqn{52}--\eqn{54}.
The generic conditions to fulfill
the requirements that leads to multiplet shortenings are thus
\bea
\label{lapri}
j=l=\left|\frac{r}{2}\right|, & H_0+4&=\left(\frac{3}{2}
r+2\right)^2  \\
\label{lasec}
\left.\begin{array}{c}
j=l-1=\left|\dfrac{r}{2}\right|\\
l=j-1=\left|\dfrac{r}{2}\right|
\end{array}\right\}, &H_0+4&=\left(\frac{3}{2}r+4\right)^2,
\eea
to which it must be added the very peculiar one
\beq
j=l=\frac{r-2}{2}, \qquad r\ge 2  
\qquad H_0 + 4 = \left( \frac{3}{2}r-2\right)^2
\eeq
which gives rise to unitary shortenings only in one case \cite{CDD}.

We have thus added some symbols in the columns at the left of the 
tables to denote the surviving states in the shortened multiplets: 
chiral ($\bullet$), semi--long ($\star$) or massless ($\diamond$) 
multiplets.

In particular, the absence of these symbols in table 4 means that no 
shortening of any kind can occur for the gravitino multiplet II.

\bigskip

As explained above, the columns regarding the $(s_1,s_2)$, $E_0$, 
$R$-symmetry values and the surviving states under shortening are 
determined by purely group theoretical arguments.
Our goal is therefore to fill in the remaining two right columns of 
all tables with the value of the masses (mass squared for all the 
fields related to second order operators), fitting unambiguously 
every 
KK field of the spectrum at the right place.

The method used to complete the multiplets is ``by exhaustion'' 
\cite{M111,CFN}: it consists in starting from the highest spin 2 
states, whose masses are unambiguously defined by the eigenvalue 
$\lambda_{[0,0]}=H_0$ and determining their energy label making use 
of 
\eqn{Delta}.
Once the $E_0$ of the multiplet (that in the graviton case belongs to 
one of the vector fields having $R$--charge $r$) is identified, one 
uses the inverses of \eqn{Delta} and `predicts' the masses of all the 
remaining states, and identifies them among the KK mass eigenvalues.
 At the end of this process, after the whole graviton multiplet has 
 been filled, part of the gravitino eigenvalues will still be 
unused.  
 As they don't match any graviton, they must generate their own spin 
 $\frac{3}{2}$ multiplets, and one repeats the above procedure of 
 arranging the lower spin states.
Again, the leftover vectors eigenmodes will start vector multiplets 
and the filling process is reiterated until all the mass values have 
been used.

Note that in the graviton multiplet and the first two families of 
vector multiplets all masses are written in terms of the usual $H_0$
\beq
H_0(j,l,r)=6 \left(j(j+1)+l(l+1)-\frac{1}{8} r^2\right),
\eeq
where $r$ is the $R$ charge of the fundamental state in the multiplet,
while
for the  gravitino multiplets there appear the shifted quantities
\beq
H_0^{\pm} \equiv H_0(j,l,r\pm 1)=H_0-\frac{3}{4}(1\pm 2r)
\eeq
and similar expressions
\beq
H_0^{\pm\pm} \equiv H_0(j,l,r\pm 2)=H_0-3(1\pm r)
\eeq
are used for the last two families of vector multiplets.
This shifts are necessary in order to match the expected
masses with formulae \eqn{acca}--\eqn{camp}, obtained from the
eigenvalues $\lambda_{[i,j]}$, where one happens to find  $H_0$ with
shifted $r$.

It is useful to list also the inverse of the formulae
\eqn{Delta}, that give directly the masses or mass squared of the 
various
$(s_1,s_2)$ representations, each having $(2 s_1+1)(2 s_2+1)$
degrees of freedom:
\beq
\label{invDelta}
\bet{crcl}
$(1,1)$                                & $m^2$  &=& $E_0 (E_0 - 4)  
$      \\
$\left(\frac{1}{2},\frac{1}{2}\right)$ & $m^2$  &=& $(E_0 - 1) (E_0 - 
3) $ \\
$(1,0)$, $(0,1)$                       & $m$    &=& $|E_0 - 
2|$             \\
$\left.
\bet{c}
$(\frac{1}{2},0)$, $(0,\frac{1}{2}),$\\
$(\frac{1}{2},1)$, $(1,\frac{1}{2})$
\eet
\right\}$                              & $m$    &=& $|E_0 - 2| $\\
$(0,0)$                                & $m^2$  &=& $E_0 (E_0 - 4) $
\eet
\eeq

For the sake of clarity, we give an example by explaining in detail
the assembling of the graviton multiplet.

The first line in table 2 is  filled without effort, since \eqn{acca}
tells that  the graviton mass spectrum is given unambiguously by the 
scalar
harmonic eigenvalue $m^2(H_{\mu\nu})\equiv H_0(j,l,r)$,
where $r$ coincides with the $R$--charge of the basic vector state 
for the
multiplet, having energy label $E^{(1)}_0=E_0$.
Since $E^{(2)}_0=E_0+1$, we can derive by \eqn{Delta} that  the 
energy label
for the whole multiplet is
\beq
\label{elab}
E_0=1+\sqrt{H_0+4}.
\eeq
Now we pass to the left and right gravitini, having generically mass 
spectrum given by \eqn{camp}
\beq
\label{speg}
m(\psi_\mu^{L,R})=\lambda_{[\frac{1}{2},\frac{1}{2}]}-\frac{5}{2} =
\left\{\begin{array}{c}
-2 \pm \sqrt{ H_0(j,l,r\pm 1)+4} \\
 -3 \pm \sqrt{H_0(j,l,r\pm1)+4}
       \end{array} \right. .
\eeq
The predicted masses  for each of the four spin $3/2$ states are 
computed by inserting the appropriate energy, given in terms of 
\eqn{elab}, and
$R$-charges in formulae \eqn{invDelta}. Due to the presence of
absolute values (and below, for spinors and scalars also of double 
sign choices), the set found in this way is redundant.
Among the eight choices of eigenmodes \eqn{speg} , only the four 
upper 
ones have a match within the predicted set of the graviton multiplet, 
while the remaining lower ones will be highest spin states for 
gravitino multiplets.

The same procedure is applied to identify the four vector states, by 
matching formulae \eqn{invDelta} with the available spin one fields 
and eigenvalues within the choices \eqn{amu}--\eqn{bifi}, according 
to 
the different energy and $R$-charges.
We point out that while the $a_\mu$ field is selected without 
ambiguity, we have chosen to assign the $B_\mu$ field with the 
positive signs in \eqn{bifi} and $\phi_\mu$ the negative ones.  The 
remaining eigenvalues will either fit in gravitino multiplets or give 
rise to vector multiplets on their own.

Analogously, the 2--form fields, spinors and the scalars are placed 
giving rise to the generic massive {\it long} graviton multiplet, 
existing for arbitrary quantum numbers $(j,l,r)$ and having an 
irrational value of $E_0$.

Then we consider the various group-theoretical shortening patterns.  
If condition \eqn{lapri} is imposed, some of the states drop out of 
the graviton multiplet that reduces to the semi--long multiplet 
identified with the $\star$ symbols.

The massless graviton multiplet identified by the $\diamond$ is 
obtained by further setting $j=l=r=0$, leading to $H_0=0$ and $E_0=3$.

The application of the above method yields the completion of all 
other 
tables, whose shortening patterns have been treated to great depth in 
\cite{CDD}, to which we refer for any detail.

\bigskip

\section{The Betti 3--form}

In this last section we want to add some considerations on the 
existence of the Betti vector multiplet in the KK spectrum previously 
discussed.
It is obvious that Betti multiplets always exist when we expand in 
harmonics a $(p+1)$--form on the higher--dimensional space considered 
as a vector on $AdS_{D-d}$ and a $p$--form on the internal compact 
space $X_d$ provided $X_d$ contains non--trivial homology $p$--cycles 
(or $(D-p)$--cycles by Poincare' duality).
Indeed, the first appearance was discussed in KK $AdS_4 \times S^7$ 
compactification \cite{DaF}, while the corresponding physical 
meaning was first discussed in \cite{Wittenbarion} where the Betti 
vector multiplets have been interpreted as topological modes 
corresponding to the wrapping of $p$--branes around the internal 
$p$--cycles.

The interesting fact emerging from the analysis we presented is that 
in the $AdS_5$ case we have not only the single Betti vector 
multiplet but also a Betti tensor and a Betti hypermultiplet deriving 
from the $A_{\mu\nu ab}$ and $A_{ab}$ fields, which are expanded in 
terms of the 2--form harmonic $Y_{[ab]}$ containing the 2--form dual 
to the Betti form.

In this section we determine the explicit form of the Betti 3--form 
on $T^{11}$ using the theorems proven in \cite{DaF,Wittold}.
They are 
\begin{enumerate} 
\item 
The Betti $p$--form is valued in the 
holonomy algebra of the internal space $X_d$; 
\item 
The Betti 
$p$--form is in the singlet representation of the isometry group 
$G$.  
\end{enumerate}

\noindent
Let us first determine the holonomy algebra on $T^{11}$.

\subsection{The holonomy algebra}

We have already shown with the harmonic analysis the existence of two 
Killing spinors for the $T^{11}$ space, implying $\cN = 2$ 
supersymmetry in the bulk.
We want now to refine this result by explicitly deriving the holonomy 
algebra $\cal H$.

A Killing spinor is a zero mode of all the holonomy algebra 
generators 
\beq
T_{\cal H} \eta = 0
\eeq
and therefore the number of preserved supersymmetries is equal to the 
number of independent solutions of the above equation.

The holonomy algebra must be a subalgebra of the tangent space one 
${\cal H}
\subset SO(5)$ and in order to have $\cN = 2$ supersymmetry $\cal H$
must be equal to $SU(2)$ \cite{Rom}, corresponding to the canonical
embedding $SO(3) \hookrightarrow SO(5)$:
\beq
\underline{5} \to \underline{3} + \underline{1} + \underline{1}.
\eeq

Let us derive this result by explicitly building the $T_{\cal H}$ 
generators annihilating the Killing spinors.
The $T_{\cH}$ generators can be constructed by the analysis of the 
integrability condition of the Killing spinor equation
\beq
\cD_a \eta = \frac{e}{2} \gamma_a \eta.
\eeq
This equation admits solutions if and only if it is integrable, i.e.
\beq
\ [\cD_a,\cD_b]\eta =  \frac{1}{4} R_{ab}{}^{cd} \gamma_{cd} \eta = 
\frac{e^2}{4}
\gamma_{ab} \eta
\eeq
which can be recast in the simpler expression
\beq
(R_{ab,cd} \gamma^{cd} - \gamma_{ab}) \eta = 0.
\eeq

It is now straightforward to implement this equation by inserting the 
Riemann curvature definitions \eqn{Riemann}.
Decomposing the antisymmetric couple $ab$ according to $a,b =
i,s,5$, we find
\bea
ij: && (3 \gamma_{12} + 2 \gamma_{34} - \gamma_{12} )\eta = 0,\\
st: && (3 \gamma_{34} + 2 \gamma_{12} - \gamma_{34} )\eta = 0,\\
i5: && 0=0,\\
s5: && 0=0,\\
is: && (\gamma_{24} - \gamma_{13} )\eta = 0, \nonumber \\
&&  (\gamma_{23} + \gamma_{14} )\eta = 0.
\eea
The three independent elements of $T_{\cH}$ (fixing
the normalisation in a convenient way) are thus given by
\bea
g_1 &=& -\frac{1}{2}(\gamma_{12} +  \gamma_{34}), \\
g_2 &=& -\frac{1}{2}(\gamma_{24} - \gamma_{13}), \\
g_3 &=& -\frac{1}{2}(\gamma_{23} + \gamma_{14}),
\eea
and using the gamma matrix algebra, one easily concludes that the
$g_i$ close the $SU(2)$ algebra
\beq
\ [g_i,g_j] = \e_{ijk} g_k.
\eeq

Recalling that the Killing spinors \eqn{eta} derived previously from
group theoretical arguments are singlet under the holonomy algebra, 
setting
\beq
\label{Omega}
h = \gamma^{abc} \Omega_{abc},
\eeq
then $h$ annihilates all the Killing spinors: $h \eta = 0$.

\subsection{The Betti 3--form}

It is known \cite{Wittold} that a harmonic 3--form $\Omega$ on the 
$T^{11}$ manifold must be in the singlet representation of the 
isometry group.  This implies that its quantum numbers are $j=l=r=0$.

Using the duality relation on the forms, the element \eqn{Omega} 
becomes
\beq
\tilde{h} = \gamma^{ab}\tilde{\Omega}_{ab}.
\eeq
Recalling the analysis of the $U_H(1)$ decomposition of the 2--form 
given in sect.  4 \eqn{vecdue} and taking into account that $h$ must 
be a singlet of the full isometry group $SU(2) \times SU(2) \times 
U_R(1)$, the only components of $\Omega_{ab}$ differing from zero are 
$\tilde{\Omega}_{+-}$ and $\tilde{\Omega}_{\hat{+}\hat{-}}$ (in 
complex notation) or $\tilde{\Omega}_{12}$ and $\tilde{\Omega}_{34}$ 
(in real notation) and that they must take real constant values 
(Since they contain only the singlet harmonic $Y_{(0)}^0$ in their 
expansion).

If we set 
\beq
\tilde{\Omega}_{+-} = a, \qquad \tilde{\Omega}_{\hat{+}\hat{-}}=b,
\qquad a,b \in \IR,
\eeq
then the condition for $h$ to be valued in the holonomy algebra $\cH$ 
reads
\beq
\tilde{h} = a \gamma^{12} + b\gamma^{34} = \a g_1 + \b g_2 + \gamma
g_3
\eeq
whose solution is $\a = a = b$ and $\b = \g = 0$ implying
that $\tilde{h} = a \, g_1$.
Thus, fixing $a=1$ for convenience, the Betti form is simply
\beq
\Omega = \star \left[ V^1 V^2 +V^3V^4\right].
\eeq

%%%%%%%%%%%%%%%%%%%%%%%%%%%%%%%%%%%
\vskip 1truecm

\paragraph{Acknowledgements.} \ We would like to thank S.  Ferrara 
for enlightening discussions on many aspects of the paper. A.C. is very
grateful to CERN for its kind hospitality during the early stages of
this work.  This research 
is supported in part by EEC under TMR contract ERBFMRX-CT96-0045.
%%%%%%%%%%%%%%%%%%%%%%%%%%%%%%%%%%%%

\section*{Appendix A: Notations and Conventions}
\setcounter{equation}{0}
\makeatletter
\@addtoreset{equation}{section}
\makeatother
\renewcommand{\theequation}{A.\arabic{equation}}

Consider $AdS_5 \times T^{11}$.
We call $M,N$ the curved ten--dimensional indices,
$\mu,\nu$/$m,n$
the curved/flat $AdS_5$ ones and  $\a\beta$/$a,b$ the curved/flat
$T^{11}$ ones.

Our ten--dimensional metric is the mostly minus $\eta = 
\{+-\ldots-\}$, so that the internal space has a negative definite 
metric.  For ease of construction, we have also used a negative 
metric 
to raise and lower the $SU(2)\times SU(2)$ Lie--algebra indices.

Furthermore, for the $SU(2)$ algebras we defined $\e^{123} = \e^{12} 
= 
1$.

The $SO(5)$ gamma matrices are
\beq
\gamma_1 = \left(
\bet{cccc}
&&&1 \\
&& 1 & \\
& -1 && \\
-1 &&&
\eet
\right) \qquad
\gamma_2 = \left(
\bet{cccc}
&&&-i \\
&& i & \\
& i && \\
-i &&&
\eet \right)
\eeq
\beq
\gamma_3 = \left(
\bet{cccc}
&&1& \\
&&&-1 \\
-1&&& \\
 &1&&
\eet \right) \qquad \gamma_4 = \left(
\bet{cccc}
&&i& \\
&&&i \\
i& && \\
 &i&&
\eet \right)
\eeq
\beq
\gamma_5 = \left(
\bet{cccc}
i&&& \\
&i&& \\
&&-i& \\
&&&-i
\eet \right)
\eeq

\section*{Appendix B: }

\setcounter{equation}{0}
\makeatletter
\@addtoreset{equation}{section}
\makeatother
\renewcommand{\theequation}{B.\arabic{equation}}

As in section 3.6, given a representation $\{\nu\}$ of $G$, we call 
the index spanning the representation space $m$ ($m = 1, \ldots, 
\hbox{dim}\{\nu\}$), while we denote by $h_i$ an index ranging in the 
subset spanned by the fragment $\{\a_i\}$ ($h_i = 1, \ldots, 
\hbox{dim}\{\a_i\}$).

Let us start from the left--invariant one--form on $G/H$ decomposed 
along the $T_H \in ${\msbm H} and $T_a \in ${\msbm K} generators 
according to the Lie algebra decomposition {\msbm G} = {\msbm H} + 
{\msbm 
K}: 
\beq
L^{-1} d L = \omega^H T_H + V^a T_a
\eeq
where $V^a$ are the vielbeins of $G/H$ and $\omega^H$ is the 
so--called $H$--connection.
We have
\beq
\label{B2}
\cD L^{-1} =  - V^a (T_a L^{-1} + \omega_a^H T_H L^{-1})
\eeq

We have by definition of harmonic $(L^{-1})^m{}_{h_i} \equiv
Y^{\{\nu\}m}_{h_i}$.
From \eqn{B2} we get
\bea
(T_a L^{-1})^m{}_{h_i} &=& (T_a)^n{}_{h_i} (L^{-1})^m{}_{n} =
\sum_{j=1}^{M} (T_a)_{h_i}{}^{h_j} Y^{\{\nu\}m}_{h_j} \\
(T_H L^{-1})^m{}_{h_i} &=& (T_H)^{k_i}{}_{h_i} (L^{-1})^m{}_{k_i} =
(T_H)_{h_i}{}^{k_i} Y^{\{\nu\}m}_{k_i}
\eea
where in the latter we have used the fact that, being $\{\a_i\}$ 
irreducible,
only the entries in the $i$--th block $(T_h)^{k_i}_{h_i}$ are
non--vanishing.

Hence, \eqn{B2} becomes (omitting the index $\{\nu\}$ denoting the
$G$--representation)
\beq
\label{B3}
dY^{m}_{h_i}=-V^a\left(\sum_{j=1}^N (T_a)_{h_i}{}^{h_j}Y^{m}_{h_i} +
 \omega_a^H (T_H)_{h_i}{}^{k_i}Y^{m}_{h_i}\right)
\eeq
or, introducing the $H$--covariant derivative
\beq
\cD^H \equiv \cD(\omega^H)
\eeq
we have
\beq
\cD^H Y^{m}_{h_i} = - r(a) V^a \sum_{j=1}^N
(T_a)_{h_i}{}^{h_j}Y^{m}_{h_i},
\eeq
where we have taken into account the vielbein rescaling $r(a)$ 
introduced in \eqn{rescaled}.
this is formula \eqn{DHH} of the text.

On the other hand, an $SO(d)$ harmonic (see equation \eqn{3199})
has $SO(d)$ covariant derivative given by
\beq
\label{B4}
\cD Y^{m}_{h_i} = d Y^{m}_{h_i}+(\cB^{cd})(T_{cd})_{h_i}{}^{k_i} 
Y^{m}_{k_i}.
\eeq
where again we have used the fact that $T_{ab}$ is block--diagonal
under the branching $SO(d) \to H$.
Decomposing $\cB^{ab}$ into the $H$--connection plus more we have
\beq
\cB^{cd}T_{cd}=\omega^H T_H+M^{cd} T_{cd}
\eeq
and substituting in \eqn{B4} we obtain
\bea
\cD Y^{m}_{h_i} &=& r(a) V^a \sum_{j=1}^N
(T_a)_{h_i}{}^{h_j} Y^{m}_{h_j} + M^{cd}(T_{cd})_{h_i}{}^{k_i} 
Y^{m}_{k_i}
\eea
which gives rise to equation \eqn{DL-1} of the text.
%%%%%%%%%%%%%%%%%%%%%%%%%%%%%%%%%%%%%%%%%%%%%%%%%%%%%%%%%%%%%%%%%%%%%%%%%%%%%%%%%%%%%%%%

\section*{Appendix C: Tables}

\begin{footnotesize}

\begin{center}

{\bf  Graviton Multiplet} \qquad $E_0=1+\sqrt{H_0+4}$.

\vskip .3cm
 \begin{tabular}{|c|c|c|c|c|c|c|}\hline
 & & $(s_1,s_2)$ &   $E_0^{(s)}$     &   $R$--symm.   &   field & 
Mass      \\
 \hline
 \hline
 $\diamond$ &$\star$ & (1,1)            & $E_0+1$      &$r$      
&$g_{\mu\nu}$
     & $H_0$               \\\hline
  $\diamond$ &$\star$ &(1,{\small 1/2})          &$E_0+1/2$     
&$r-1$    &$\psi_\mu^L$         &$-2+\sqrt{H_0+4}$     \\
  $\diamond$ &$\star$ &(1/2,1)          &$E_0+1/2$     &$r+1$    
&$\psi_\mu^R$
         &$-2+\sqrt{H_0+4}$     \\
   &$\star$ &(1/2,1)          &$E_0+3/2$     &$r-1$    &$\psi_\mu^R$
     &$-2-\sqrt{H_0+4}$   \\
   & &(1,1/2)          &$E_0+3/2$     &$r+1$    &$\psi_\mu^L$
&$-2-\sqrt{H_0+4}$      \\
 \hline
  $\diamond$ &$\star$ &(1/2,1/2)     &$E_0$  &$r$    &$ \phi_\mu$ 
&$H_0+4-2\sqrt{H_0+4}$    \\
   & &(1/2,1/2)     &$E_0+1$&$r+2$  &$ a_\mu$  &$H_0+3$  \\
   &$\star$ &(1/2,1/2)     &$E_0+1$&$r-2$  &$ a_\mu$   & $H_0+3$  \\
   & &(1/2,1/2)     &$E_0+2$&$r$    &$B_\mu$   
&$H_0+4+2\sqrt{H_0+4}$   \\
\hline
   & &(1,0)     & $E_0+1$      &$r$  &$b_{\mu\nu}^+$ 
&$\sqrt{H_0+4}$    \\
   &$\star$ &(0,1)     & $E_0+1$      &$r$  &$b_{\mu\nu}^-$ 
&$-\sqrt{H_0+4}$     \\
\hline
   & &(1/2,0) & $E_0+1/2$      &$r+1$  & 
$\lambda_L$&$1/2-\sqrt{H_0+4}$  \\
   &$\star$ &( 0,1/2)& $E_0+1/2$      &$r-1$  & 
$\lambda_R$&$1/2-\sqrt{H_0+4}$   \\
   & &(1/2,0)  & $E_0+3/2$     &$r-1$  & 
$\lambda_L$&$1/2+\sqrt{H_0+4}$    \\
   & &(0,1/2)  & $E_0+3/2$ &$r+1$  &$\lambda_R$  &$1/2+\sqrt{H_0+4}$  
\\
\hline
   & &(0,0)  & $E_0+1$&$r$ & $ B$ & $H_0$     \\
\hline
 \end{tabular}

%%%%%%%%%%%%%%%%%

\vspace{1cm}

{\bf Gravitino Multiplet I}\qquad $E_0=\sqrt{H_0^-+4}-1/2$

\vskip .3cm
 \begin{tabular}{|c|c|c|c|c|c|c|}\hline
 & & $(s_1,s_2)$ &   $E_0^{(s)}$     &   $R$--symm.   &   field & 
Mass      \\
 \hline \hline
&$\star$ & (1,1/2) &$E_0+1$&$r$&$\psi_\mu^L$&$-3+\sqrt{H_0^-+4}$\\
 \hline
   &$\star$ &(1/2,1/2)          &$E_0+1/2$     &$r+1$    
&$\phi_\mu$         &$H_0^-+7-4\sqrt{H_0^-+4}$  \\
   &$\star$ &(1/2,1/2)          &$E_0+3/2$     &$r-1$    &$ 
a_\mu$         &$H_0^-+4-2\sqrt{H_0^-+4}$     \\
 \hline
  $\bullet$ &$\star$ &(1,0)          &$E_0+1/2$     &$r-1$    &$ 
a_{\mu\nu}$
  &$2-\sqrt{H_0^-+4}$  \\
   & &(1,0)        &$E_0+3/2$     &$r+1$    &$b_{\mu\nu}^+$         
&$1-\sqrt{H_0^-+4}$      \\
 \hline
  $\bullet$ &$\star$ &(1/2,0)     &$E_0$  &$r$    &$ \psi^{(T)}_L$ 
&$-5/2+\sqrt{H_0^-+4}$    \\
   $\bullet$&$\star$ &(1/2,0)     &$E_0+1$&$r-2$  &$\psi^{(T)}_L$  
&$-3/2+\sqrt{H_0^-+4}$  \\
  & $\star$ &(0,1/2)     &$E_0+1$&$r$  &$\lambda_R$   & 
$3/2-\sqrt{H_0^-+4}$  \\
  &  &(1/2,0)     &$E_0+1$&$r+2$    &$\psi^{(T)}_L$   
&$-3/2+\sqrt{H_0^-+4}$   \\
 & &(1/2,0)     &$E_0+2$      &$r$  &$\psi^{(T)}_L$ 
&$-1/2+\sqrt{H_0^-+4}$    \\
 \hline
   $\bullet$&$\star$ &(0,0)     & $E_0+1/2$      &$r-1$  &$a$ 
&$H_0^-+4-4\sqrt{H_0^-+4}$    \\
 & &(0,0)     & $E_0+3/2$      &$r+1$  & 
$a$&$H_0^-+1-2\sqrt{H_0^-+4}$  \\
\hline
 \end{tabular}

\newpage

{\bf  Gravitino Multiplet II}\qquad $E_0=5/2 +\sqrt{H_0^++4}$
\vskip .3cm
 \begin{tabular}{|c|c|c|c|c|}\hline
   $(s_1,s_2)$ &   $E_0^{(s)}$     &   $R$--symm.   &   field & 
Mass      \\
 \hline
 \hline
(1,1/2)&$E_0+1$&$r$&$\psi_\mu^L$&$-3-\sqrt{H_0^++4}$\\
 \hline
(1/2,1/2)          &$E_0+1/2$     &$r+1$    &$a_\mu$         
&$H_0^++4+2\sqrt{H_0^++4}$     \\
(1/2,1/2)          &$E_0+3/2$     &$r-1$    &$B_\mu$         
&$H_0^++7+4\sqrt{H_0^++4}$\\
 \hline
(1,0)          &$E_0+1/2$     &$r-1$    &$b_{\mu\nu}^+$         
&$1+\sqrt{H_0^++4}$      \\
(1,0)        &$E_0+3/2$     &$r+1$   &$a_{\mu\nu}$         
&$2+\sqrt{H_0^++4}$  \\
\hline
(1/2,0)     &$E_0$  &$r$    &$ \psi^{(T)}_L$ 
&$-1/2-\sqrt{H_0^++4}$    \\
(1/2,0)     &$E_0+1$&$r-2$  &$\psi^{(T)}_L$  &$-3/2-\sqrt{H_0^++4}$  
\\
(0,1/2)     &$E_0+1$&$r$  &$\lambda_R$   & $3/2+\sqrt{H_0^++4}$  \\
(1/2,0)     &$E_0+1$&$r+2$    &$\psi^{(T)}_L$   
&$-3/2-\sqrt{H_0^++4}$   \\
(1/2,0)     &$E_0+2$      &$r$  &$\psi^{(T)}_L$ 
&$-5/2-\sqrt{H_0^++4}$    \\
 \hline
(0,0)     & $E_0+1/2$      &$r-1$  &$a$ &$H_0^++1+2\sqrt{H_0^++4}$    
\\
(0,0)     & $E_0+3/2$      &$r+1$  & $a$&$H_0^++4+4\sqrt{H_0^++4}$  \\
\hline
 \end{tabular}

%%%%%%%%%%%%%%%%%

\vspace{1cm}

{\bf  Gravitino Multiplet III}\qquad $E_0=-1/2+\sqrt{H_0^++4}$
\vskip .3cm
 \begin{tabular}{|c|c|c|c|c|c|}\hline
  & $(s_1,s_2)$ &   $E_0^{(s)}$     &   $R$--symm.   &   field & 
Mass      \\
 \hline \hline
$\star$ & (1/2,1) &$E_0+1$&$r$&$\psi_\mu^R$&$-3+\sqrt{H_0^++4}$\\
 \hline
   $\star$ &(1/2,1/2)          &$E_0+1/2$     &$r-1$    
&$\phi_\mu$         &$H_0^++7-4\sqrt{H_0^++4}$  \\
    &(1/2,1/2)          &$E_0+3/2$     &$r+1$    &$ a_\mu$         
&$H_0^++4-2\sqrt{H_0^++4}$     \\
 \hline
  $\star$ &(0,1)          &$E_0+1/2$     &$r+1$    &$ 
a_{\mu\nu}$         &$2-\sqrt{H_0^++4}$  \\
   $\star$ &(0,1)        &$E_0+3/2$     &$r-1$    
&$b_{\mu\nu}^-$         &$1-\sqrt{H_0^++4}$      \\
 \hline
  $\star$ &(0,1/2)     &$E_0$  &$r$    &$ \psi^{(T)}_R$ 
&$-5/2+\sqrt{H_0^++4}$    \\
   $\star$ &(0,1/2)     &$E_0+1$&$r+2$  &$\psi^{(T)}_R$  
&$-3/2+\sqrt{H_0^++4}$  \\
    &(1/2,0)     &$E_0+1$&$r$  &$\lambda_L$   & $3/2-\sqrt{H_0^++4}$  
\\
    &(0,1/2)     &$E_0+1$&$r-2$    &$\psi^{(T)}_R$   
&$-3/2+\sqrt{H_0^++4}$   \\
    &(0,1/2)     &$E_0+2$      &$r$  &$\psi^{(T)}_R$ 
&$-1/2+\sqrt{H_0^++4}$    \\
 \hline
    &(0,0)     & $E_0+1/2$      &$r+1$  &$a$ 
&$H_0^++4-4\sqrt{H_0^++4}$    \\
    &(0,0)     & $E_0+3/2$      &$r-1$  & 
$a$&$H_0^++1-2\sqrt{H_0^++4}$  \\
\hline
 \end{tabular}

\newpage

{\bf  Gravitino Multiplet IV}\qquad $E_0=5/2+\sqrt{H_0^-+4}$

\vskip .3cm
 \begin{tabular}{|c|c|c|c|c|c|}\hline
  & $(s_1,s_2)$ &   $E_0^{(s)}$     &   $R$--symm.   &   field & 
Mass      \\
 \hline
 \hline
$\star$&(1/2,1)&$E_0+1$&$r$&$\psi_\mu^R$&$-3-\sqrt{H_0^-+4}$\\
 \hline
$\star$&(1/2,1/2)          &$E_0+1/2$     &$r-1$    &$a_\mu$         
&$H_0^-+4+2\sqrt{H_0^-+4}$     \\
&(1/2,1/2)          &$E_0+3/2$     &$r+1$    &$B_\mu$         
&$H_0^-+7+4\sqrt{H_0^-+4}$\\
 \hline
$\star$&(0,1)          &$E_0+1/2$     &$r+1$    
&$b_{\mu\nu}^-$         &$1+\sqrt{H_0^-+4}$      \\
$\star$&(0,1)        &$E_0+3/2$     &$r-1$   &$a_{\mu\nu}$         
&$2+\sqrt{H_0^-+4}$  \\
 \hline
$\star$&(0,1/2)     &$E_0$  &$r$    &$ \psi^{(T)}_R$ 
&$-1/2-\sqrt{H_0^-+4}$    \\
$\star$&(0,1/2)     &$E_0+1$&$r+2$  &$\psi^{(T)}_R$  
&$-3/2-\sqrt{H_0^-+4}$  \\
&(1/2,0)     &$E_0+1$&$r$  &$\lambda_L$   & $3/2+\sqrt{H_0^-+4}$  \\
&(0,1/2)     &$E_0+1$&$r-2$    &$\psi^{(T)}_R$   
&$-3/2-\sqrt{H_0^-+4}$   \\
&(0,1/2)     &$E_0+2$      &$r$  &$\psi^{(T)}_R$ 
&$-5/2-\sqrt{H_0^-+4}$    \\
 \hline
&(0,0)     & $E_0+1/2$      &$r+1$  &$a$ 
&$H_0^-+1+2\sqrt{H_0^-+4}$    \\
&(0,0)     & $E_0+3/2$      &$r-1$  & $a$&$H_0^-+4+4\sqrt{H_0^-+4}$  
\\
\hline
 \end{tabular}

%%%%%%%%%%%%%%%%%

\bigskip

{\bf Vector Multiplet I} \qquad $E_0=\sqrt{H_0+4}-2$

\vskip .3cm
 \begin{tabular}{|c|c|c|c|c|c|c|c|}\hline
&   &  & $(s_1,s_2)$    & $E_0^{(s)}$ &   $R$--symm.   &   field & 
Mass      \\
 \hline
 \hline
$\diamond$&&$\star$&(1/2,1/2)            &$E_0+1$      &$r$
&$\phi_\mu$&$H_0+12-6\sqrt{H_0+4}$\\
 \hline
$\diamond$& $ \bullet$ & $\star$ &(1/2,0)          &$E_0+1/2$     
&$r-1$    &$\psi^{(L)}_L$         &$7/2-\sqrt{H_0+4}$  \\
$\diamond$&   &$\star$  &(0,1/2)                  &$E_0+1/2$     
&$r+1$    &$\psi^{(L)}_R$         &$7/2-\sqrt{H_0+4}$     \\
  & &$\star$  &(0,1/2)          &$E_0+3/2$     &$r-1$    
&$\psi^{(L)}_R$         &$5/2-\sqrt{H_0+4}$  \\
  & & &(1/2,0)&$E_0+3/2$     &$r+1$    
&$\psi^{(L)}_L$         
&$5/2-\sqrt{H_0+4}$      \\
 \hline
 $\diamond$& $\bullet$ &$\star$ &(0,0)     &$E_0$  &$r$    &$b$ 
&$H_0+16-8\sqrt{H_0+4}$    \\
  & $\bullet$&$\star$ &(0,0)     &$E_0+1$&$r-2$  &$\phi$  
&$H_0+9-6\sqrt{H_0+4}$  \\
  & & &(0,0)     &$E_0+1$&$r+2$  &$\phi$   & $H_0+9-6\sqrt{H_0+4}$  \\
  & & &(0,0)     &$E_0+2$&$r$    &$\phi$   &$H_0+4-4\sqrt{H_0+4}$   \\
 \hline
 \end{tabular}

\bigskip

{\bf Vector Multiplet II} \qquad $E_0=\sqrt{H_0+4}+4$

\vskip .3cm
 \begin{tabular}{|c|c|c|c|c|}\hline
      $(s_1,s_2)$   & $E_0^{(s)}$ &   $R$--symm.   &   field & 
Mass      \\
 \hline
 \hline
(1/2,1/2)            &$E_0+1$      &$r $&$B_\mu  
$&$H_0+12+6\sqrt{H_0+4}$\\
 \hline
   (1/2,0)        &$E_0+1/2$     &$r-1$    
&$\psi^{(L)}_L$         &$5/2+\sqrt{H_0+4}$  \\
   (0,1/2)        &$E_0+1/2$     &$r+1$    
&$\psi^{(L)}_R$         &$5/2+\sqrt{H_0+4}$     \\
  (0,1/2)        &$E_0+3/2$     &$r-1$    
&$\psi^{(L)}_R$         &$7/2+\sqrt{H_0+4}$  \\
    (1/2,0)        &$E_0+3/2$     &$r+1$    &$\psi^{(L)}_L$         
&$7/2+\sqrt{H_0+4}$      \\
 \hline
  (0,0)     &$E_0$  &$r$    &$\phi$ 
&$H_0+4+4\sqrt{H_0+4}$    \\
   (0,0)     &$E_0+1$&$r-2$  &$\phi$  &$H_0+9+6\sqrt{H_0+4}$  
\\
    (0,0)     &$E_0+1$&$r+2$  &$\phi$   & $H_0+9+6\sqrt{H_0+4}$  \\
    (0,0)     &$E_0+2$&$r$    &$\pi$   &$H_0+16+8\sqrt{H_0+4}$   \\
 \hline
 \end{tabular}

\newpage

{\bf Vector Multiplet III} \qquad $E_0=\sqrt{H_0^{++}+4}+1$;

\vskip .3cm
 \begin{tabular}{|c|c|c|c|c|c|}\hline
     & $(s_1,s_2)$    & $E_0^{(s)}$ &   $R$--symm.   &   field & 
Mass      \\
 \hline
 \hline
&(1/2,1/2)            &$E_0+1$      &$r$       &$ 
a_\mu$&$H_0^{++}+3$\\
 \hline
 &(1/2,0)          &$E_0+1/2$     &$r-1$    
&$\psi^{(T)}_L$        &$-1/2+\sqrt{H_0^{++}+4}$  \\
   &(0,1/2)                  &$E_0+1/2$     &$r+1$    
&$\psi^{(T)}_R$         &$-1/2+\sqrt{H_0^{++}+4}$     \\
   &(0,1/2)          &$E_0+3/2$     &$r-1$    
&$\psi^{(T)}_R$         &$1/2+\sqrt{H_0^{++}+4}$  \\
$\bullet$     &(1/2,0)        &$E_0+3/2$     &$r+1$    
&$\psi^{(T)}_L$         
&$1/2+\sqrt{H_0^{++}+4}$      \\
 \hline
   &(0,0)     &$E_0$  &$r$    &$a$ 
&$H_0^{++}+1-2\sqrt{H_0^{++}+4}$    \\
   &(0,0)     &$E_0+1$&$r-2$  &$\phi$  &$H_0^{++}$  \\
 $\bullet$   &(0,0)     &$E_0+1$&$r+2$  &$\phi$   & $H_0^{++}$  \\
  $\bullet$   &(0,0)     &$E_0+2$&$r$    &$a$   
&$H_0^{++}+1+2\sqrt{H_0^{++}+4}$   \\
 \hline
 \end{tabular}

\bigskip

{\bf Vector Multiplet IV} \qquad $E_0=\sqrt{H_0^{--}+4}+1$

\vskip .3cm
 \begin{tabular}{|c|c|c|c|c|c|c|}\hline
 &  &    $(s_1,s_2)$   & $E_0^{(s)}$ &   $R$--symm.   &   field & 
Mass      \\
 \hline
 \hline
 &$\star$ &(1/2,1/2)            &$E_0+1$      &$r$               
&$a_\mu$&$H_0^{--}+3$\\
 \hline
  $\bullet$ &$\star$ &(1/2,0)       &$E_0+1/2$     &$r-1$   
&$\psi^{(T)}_L$          &$-1/2+\sqrt{H_0^{--}+4}$  \\
   &$\star$ &(0,1/2)       &$E_0+1/2$     &$r+1$    
&$\psi^{(T)}_R$               &$-1/2+\sqrt{H_0^{--}+4}$     \\
   &$\star$ &(0,1/2)       &$E_0+3/2$     &$r-1$   
&$\psi^{(T)}_R$          &$1/2+\sqrt{H_0^{--}+4}$  \\
 &&(1/2,0)       &$E_0+3/2$     &$r+1$    
&$\psi^{(T)}_L$          &$1/2+\sqrt{H_0^{--}+4}$      \\
 \hline
 $\bullet$  &$\star$ &(0,0)     &$E_0$  &$r$    &$a$ 
&$H_0^{--}+1-2\sqrt{H_0^{--}+4}$    \\
 $\bullet$ &$\star$  &(0,0)     &$E_0+1$&$r-2$  &$B$  &$H_0^{--}$  \\
& &(0,0)     &$E_0+1$&$r+2$  &$\phi$   & $H_0^{--}$  \\
 & &(0,0)     &$E_0+2$&$r$    &$a$   
&$H_0^{--}+1+2\sqrt{H_0^{--}+4}$   \\
 \hline
 \end{tabular}
%%%%%%%%%%%%%%%%%%%%%%%%%%

\bigskip

\end{center}
\end{footnotesize}

%

% ---- Bibliography ----

%

\end{document}